**Approximating the Sum of Correlated Lognormals: An Implementation[†]**

By CHRISTOPHER J. ROOK[1] and MITCHELL C. KERMAN[2]

## ABSTRACT

Lognormal random variables appear naturally in many engineering disciplines, including wireless communications, reliability theory, and finance. So, too, does the sum of (correlated) lognormal random variables. Unfortunately, no closed form probability distribution exists for such a sum, and it requires approximation. Some approximation methods date back over 80 years and most take one of two approaches, either: 1) an approximate probability distribution is derived mathematically, or 2) the sum is approximated by a single lognormal random variable. In this research, we take the latter approach and review a fairly recent approximation procedure proposed by Mehta, Wu, Molisch, and Zhang (2007), then implement it using C++. The result is applied to a discrete time model commonly encountered within the field of financial economics.

[1] Christopher J. Rook works as a consultant statistical programmer and is finishing a degree in Systems Engineering from Stevens Institute of Technology.

[2] Dr. Mitchell C. Kerman is the Director of Program Development and Transition for the Systems Engineering Research Center, a Department of Defense (DoD) University Affiliated Research Center (UARC) led by Stevens Institute of Technology.

## I. Introduction

Practical problems involving sums of random variables (RVs), say, $Z = X + Y$, are unavoidable within many disciplines. When X and Y are independent, the probability density function (PDF) of Z can be expressed using the convolution operator. Much theory on RV sums has been developed and closed form expressions for the PDF of Z exist for some X and Y. In general, analytical solutions involving convolutions for sums of independent RVs are often difficult to obtain. When X and Y are correlated, the complexity increases. Sums of n independent and identically distributed (*iid*) RVs, say, $Z_n = X_1 + X_2 + \ldots + X_n$, can be approximated by the normal distribution using the central limit theorem (CLT), but only when $n \geq 30$. Therefore, if $n < 30$, or if the RVs are not *iid*, then the CLT does not apply. When $n < 30$ and the RVs are independent, we may be able to derive the PDF of $Z_n$ by applying the convolution operator iteratively. That is, we first determine the PDF of $Z_2 = X_1 + X_2$, then recognize that $Z_3 = Z_2 + X_3$ is also a 2-term sum of independent RVs, with the PDF of $Z_2$ and $X_3$ perhaps known. While theoretically sound, it may be unlikely that this technique will produce successive closed form PDFs for each sum.

Lognormal RVs appear in many disciplines including finance, fiber optics, inventory management, telecommunications, and reliability theory. By definition, a variable is said to be lognormally distributed when its logarithm is normally distributed. Lognormal RVs tend to appear naturally when a phenomenon involves the product of *iid* RVs. To see why, we can take the logarithm of the product, which becomes a sum of *iid* logged RVs that tends to the normal distribution as the number of RVs multiplied increases (via the CLT). Exponentiating the sum then yields a lognormal RV. For example, let $P_n = X_1 \cdot X_2 \cdot \cdots \cdot X_n$, where the $X_i$'s are *iid* RVs, so that $\ln(P_n) = \ln(X_1 \cdot X_2 \cdot \cdots \cdot X_n) = \ln(X_1) + \ln(X_2) + \cdots + \ln(X_n) \overset{\cdot}{\sim} N(\mu, \sigma^2)$, for $n \geq 30$. By exponentiating both sides, $P_n \overset{\cdot}{\sim} e^{N(\mu, \sigma^2)}$, that is, $P_n$ is approximately lognormal since its logarithm is approximately normal. Since lognormal RVs appear naturally in such settings, so too will their sum. Unfortunately, the convolution for the sum of two independent lognormal RVs does not have a closed form, and, therefore, neither does the PDF. Mehta, Wu, Molisch, and Zhang (2007) propose a novel and flexible approach to approximate the distribution of a sum of (correlated) lognormal RVs and in this research we review, then implement it, using C++.

The remainder of this research is organized as follows. In Section II, we review the literature on lognormal sum approximations, and, in particular, the two methods that motivate the technique proposed by Mehta et al. (2007). In Section III, we present a detailed review of the theoretical concepts required to apply the technique. This section may be skipped by readers already familiar with these concepts. In Section IV, we implement the technique for a 2-term sum, and, in Section V, we present an application to finance. In Section VI, we discuss the extension to a sum of more than two terms. Section VII concludes with our recommendations. Fully documented source code for a C++ implementation is provided in Appendices A and B.[1]

## II. Literature Review

Fenton (1960) proposed a moment-matching technique to approximate the distribution of a sum of independent lognormal RVs with a single lognormal RV. Probabilities in the center of the distribution can be approximated effectively using this method by matching the 1st and 2nd central moments (i.e., the mean and variance). If interest is in upper tail probabilities, then the approximation is derived by matching the 2nd and 3rd central moments. For values further in the tail, the approximation can be based on the 3rd and 4th central moments. This technique is not customizable for probabilities in the head portion (i.e., near zero) because a practical formula for deriving negative moments of the sum is not known. This approach was used by R.I. Wilkinson at Bell Labs in the 1930's and is therefore often referred to as the Fenton-Wilkinson (F-W) procedure. Schwartz and Yeh (1982) proposed a similar moment-matching technique but on the log scale. The procedure is iterative, handling two terms at a time and it assumes that each successive sum is lognormal, thus the log is normally distributed. Analytical expressions are provided for the 2-term case, which is sufficient to implement the procedure. Schwartz and Yeh (S-Y) report an improvement over the F-W method and show how the procedure can be applied to correlated lognormal RVs. Mehta et al. (2007) note that F-W is more accurate in the tail while S-Y is more accurate in the head of the distribution and propose a method that is customizable, allowing the user to parameterize the procedure to meet their needs. Instead of matching moments, Mehta et al. (2007) propose matching the moment-generating function (MGF) directly, and, similar to F-W and S-Y, approximate a sum of lognormals with a single lognormal RV.

---

## III. Preliminaries

In this section, we provide a general foundation for the techniques that will be involved in approximating the distribution of the sum of (correlated) lognormal RVs with a single lognormal RV. Any reader who is familiar with these topics may skip this section.

### A. The Moment-Generating Function

The MGF for a continuous RV, X, with PDF, *f(x)*, is defined by the following function of both *t* and *x* (Freund [1992]):

$$M_X(t) = E_X[e^{tx}] = \int_{-\infty}^{\infty} e^{tx} * f(x)dx \ . \tag{1}$$

In words, it is the expected value of $e^{tx}$ with respect to the RV X. For a discrete RV, we replace the integral by a sum. We refer to this as the MGF because it can be used to derive the moments of X.

The $n^{th}$ moment of X is the expected value of $X^n$, denoted $E[X^n]$. If $\mu$ and $\sigma^2$ are the mean and variance of X, they are related to the $1^{st}$ and $2^{nd}$ moments as follows:

$$\mu = E[X^1] \quad \text{and} \quad \sigma^2 = E[X^2] - (E[X^1])^2 \ . \tag{2}$$

The $n^{th}$ moment, as defined here, is sometimes referred to as a moment *about the origin* to distinguish it from moments about the mean, which are $E[(X-\mu)^n]$. To derive the $n^{th}$ moment of X using its MGF, we differentiate $M_X(t)$ n times with respect to t, then set t equal to zero (Freund [1992]). That is,

$$E[X^n] = \frac{d^n}{dt^n}[M_X(t)]\Big|_{t=0} \ . \tag{3}$$

Clearly, from this definition, it immediately follows that the MGF for the sum of two independent RVs $X_1$ and $X_2$, namely $Z = X_1 + X_2$, is the product of their respective MGFs. In this research, however, we do not assume that $X_1$ and $X_2$ are independent. Therefore, this simplified expression is of limited use.

## B. An Overview of Gaussian Quadrature

We can approximate the area under a curve (i.e., an integral) by dividing the area into rectangles of equal width and summing their areas as depicted in Figure 1 below. In place of rectangles, a more accurate estimate may be derived using trapezoids or polynomials at the top. These methods are known as the *trapezoidal rule* and *Simpson's rule*, respectively (Anton [1988]).

**Figure 1**
**Approximating the Area Under a Curve using Rectangles**

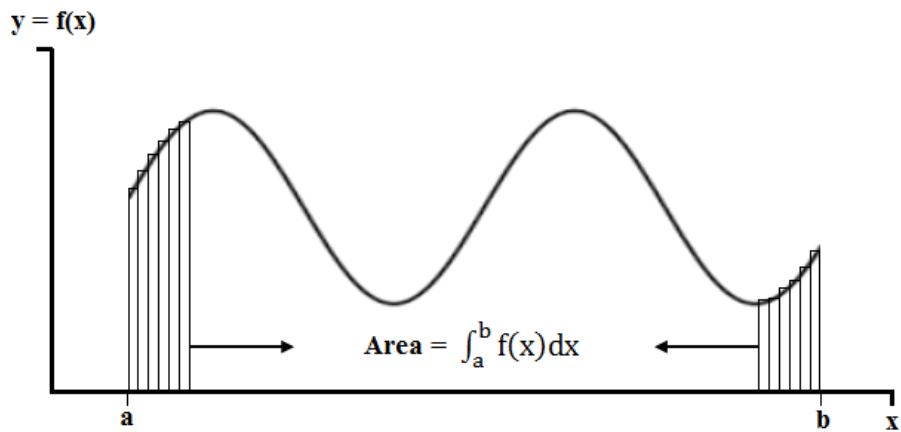

If a total of n rectangles are used in the approximation, then the width can be fixed at $w_r = \frac{b-a}{n}$, for r = 1, 2, …, n. Using a fixed width, the midpoint of the $r^{th}$ rectangle is $m_r^* = a + (r-1) * w_r + \left(\frac{w_r}{2}\right) = a + w_r * (r - \frac{1}{2})$, for r = 1, 2, …, n. The height of each rectangle is the function evaluated at $m_r^*$, namely, $f(m_r^*)$. Therefore, the area under f(x) between points a and b is estimated as:

$$\text{Area Under Curve} = \int_a^b f(x) \, dx \approx \sum_{r=1}^n w_r * f(m_r^*), \tag{4}$$

and,

$$\lim_{n \to \infty} \left( \sum_{r=1}^n w_r * f(m_r^*) \right) = \int_a^b f(x) \, dx. \tag{5}$$

The term in (5) is a Riemann sum and merely reflects the fact that using an infinite number of rectangles will yield the area exactly without it being an approximation (Anton [1988]). Depending on the function being integrated, we may need thousands of rectangles (or more) to obtain a good estimate. Clearly, if this integration appears within a larger iterative routine, then it will suffer runtime inefficiencies. A faster approximation can be achieved using numerical quadrature. This technique often requires only a small number of areas be summed, where both the rectangle "widths" and "midpoints" are determined mathematically. To estimate the area under $f(x)$ between $a$ and $b$ using numerical quadrature, we first express $f(x) = w(x)*g(x)$ where $w(x) \geq 0$ is referred to as a *weight* function (Golub and Welsch [1969]). Note that, in general, $g(x) = [w(x)]^{-1}*f(x)$. Then:

$$\int_a^b f(x)\,dx = \int_a^b w(x) * g(x)\,dx \approx \sum_{j=1}^n w_j * g(t_j)\,, \qquad (6)$$

where the pairs $(w_j, t_j)$ for $j=1, 2, \ldots, n$ have been specifically derived for $w(x)$ with respect to a given set of orthogonal polynomials. A set of polynomials $P_j(x)$ of degree $j$ for $j = 1, 2, \ldots, n+1$ are said to be orthogonal with respect to $w(x)$ if the following condition holds (Golub and Welsch [1969]):

$$\int_a^b w(x) * P_h(x) * P_k(x)\,dx = 0, \qquad \text{for } h \neq k\,. \qquad (7)$$

This strict condition requires that the set of polynomials, $P_j(x)$, be carefully constructed to satisfy it. For certain weight functions the polynomials have already been derived. The sequence of orthogonal polynomials will take the form (Golub and Welsch [1969]):

$$P_1(x) = k_1*(x - t_1)$$
$$P_2(x) = k_2*(x - t_1)*(x - t_2)$$
$$\vdots \qquad (8)$$
$$P_n(x) = k_n*(x - t_1)*(x - t_2)*(x - t_3)* \ldots *(x - t_n)$$
$$P_{n+1}(x) = k_{n+1}*(x - t_1)*(x - t_2)*(x - t_3)* \ldots *(x - t_{n+1})\,,$$

which reveals an important quality. Namely, the values $t_j$ alluded to in (6) are the roots of $P_n(x)$ and satisfy $a < t_j < b$, $\forall\, j = 1, 2, \ldots, n$. The weights, $w_j$, are calculated as functions of the polynomials and their derivatives evaluated at the roots. Once the weights, $w_j$, and roots, $t_j$, are known, the sum in (6) can be calculated to estimate the integral. This technique is referred to as numerical quadrature. In Gaussian quadrature, the formula to calculate the weights is given by (Golub and Welsch [1969]):

$$w_j = -\frac{k_{n+1}}{k_n} * \frac{1}{P_{n+1}(t_j) * P_n{}'(t_j)} \; . \tag{9}$$

We cannot provide details about the roots, $t_j$, because the orthogonal polynomials will be specific to the weight function, $w(x)$, and our only requirement at this stage is that it be greater than or equal to zero for all $x \in (a, b)$.

### C. Gauss-Hermite Quadrature

When the weight function is $w(x) = e^{-x^2}$ and $(a, b) = (-\infty, +\infty)$, the Hermite polynomials are orthogonal. That is, they satisfy (7) (Abramowitz and Stegun [1964]). Weight and root pairs $(w_j, t_j)$ have already been calculated for several n and, in this research, we use n=12 "rectangles" to approximate the necessary integrals as suggested by Mehta et al. (2007). The corresponding weights and roots are shown below in Table 1 (Abramowitz and Stegun [1964]).

**Table 1**
**Gauss-Hermite Quadrature Weights and Roots for n=12**

| | Roots ($t_j$) | Weights ($W_j$) |
|---|---|---|
| (+/-) | 0.314240376254 | 0.570135236262500000 |
| (+/-) | 0.947788391240 | 0.260492310264200000 |
| (+/-) | 1.597682635153 | 0.051607985615880000 |
| (+/-) | 2.279507080501 | 0.003905390584629000 |
| (+/-) | 3.020637025121 | 0.000085736870435880 |
| (+/-) | 3.889724897870 | 0.000000265855168436 |

### D. Overview of Lognormal RVs

In this section we present various results for lognormal RVs. Once finished, we will apply the routine suggested by Mehta et al. (2007) to approximate the sum of correlated lognormal RVs with a single new lognormal RV.

### D.1 Standard Univariate Form

Let $X \sim N(\mu_x, \sigma_x{}^2)$ be a normally distributed RV. The PDF of X, *f(x)*, is given by (Casella and Berger [1990]):

$$f(x) = \frac{1}{\sqrt{2\pi}\sigma_x} e^{-\frac{1}{2}\left(\frac{x-\mu_x}{\sigma_x}\right)^2}, \quad for -\infty < x < \infty. \tag{10}$$

The RV $Y = e^X$ is then said to be lognormally distributed. We derive the PDF of Y by making the substitution $X = \ln(Y)$ in (10) above. The Jacobian of this transformation is $\frac{dX}{dY} = \frac{1}{Y}$ and Y > 0. Therefore, the PDF of the lognormal RV Y is given by *f(y)*, where:

$$f(y) = \frac{1}{y\sqrt{2\pi}\sigma_x} e^{-\frac{1}{2}\left(\frac{\ln(y)-\mu_x}{\sigma_x}\right)^2}, \quad for\ y \geq 0. \tag{11}$$

The mean and variance of Y are $E(Y) = e^{\mu_x + \frac{\sigma_x^2}{2}}$ and $V(Y) = \left(e^{2\mu_x + \sigma_x^2}\right)\left(e^{\sigma_x^2} - 1\right)$, respectively (Walpole, Myers, Myers, and Ye [2002]).

### D.2 Alternative Univariate Form

Using the natural log to define a lognormal RV is not required and any base logarithm will suffice. In this research, base 10 will be used with a constant factor to align with Mehta et al. (2007). Namely, we will define $\ddot{Y} = 10^{X/10}$. Let $\theta = \left(\frac{\ln(10)}{10}\right)$, then $\ln(\ddot{Y}) = \theta X \rightarrow \ddot{Y} = e^{\theta X}$. Clearly, $\theta X \sim N(\theta\mu_x, \theta^2\sigma_x^2)$, therefore $\ddot{Y}$ has the standard form lognormal distribution (see Section III.D.1) based on the underlying normally distributed RV $\theta X$. Using (11) above, it immediately follows that the mean and variance of $\ddot{Y}$ are, respectively:

$$E[\ddot{Y}] = e^{\left(\theta\mu_x + \frac{\theta^2\sigma_x^2}{2}\right)} \tag{12}$$

and

$$V[\ddot{Y}] = \left(e^{2\theta\mu_x + \theta^2\sigma_x^2}\right)\left(e^{\theta^2\sigma_x^2} - 1\right). \tag{13}$$

Note that $\frac{dX}{d\ddot{Y}} = \frac{1}{\theta}\frac{1}{\ddot{Y}}$ and $\ddot{Y} > 0$. If we are given $E[\ddot{Y}]$ and $V[\ddot{Y}]$, then $E[X] = \mu_x$ and $V[X] = \sigma_x^2$ can be calculated and vice-versa (see Section III.D.4 below). Finally, using (11) with the updated underlying normal distribution, the PDF of $\ddot{Y}$, *f(ÿ)*, is given by:

$$f(\ddot{y}) = \frac{1}{\theta \ddot{y} \sqrt{2\pi} \sigma_x} e^{-\frac{1}{2}\left(\frac{\left(\frac{1}{\theta}\right)\ln(\ddot{y}) - \mu_x}{\sigma_x}\right)^2}, \quad for \; \ddot{y} \geq 0. \tag{14}$$

### D.3 Standard Bivariate Form

Let $X_1 \sim N(\mu_{x_1}, \sigma_{x_1}^2)$ and $X_2 \sim N(\mu_{x_2}, \sigma_{x_2}^2)$ be joint normal RVs with $Cov(X_1, X_2) = \sigma_{(x_1, x_2)}$. The bivariate PDF of $X_1$ and $X_2$, $f(x_1, x_2)$, is given by (Casella and Berger [1990]):

$$f(x_1, x_2) = \frac{1}{2\pi\sigma_{x_1}\sigma_{x_2}\sqrt{1-\rho^2}} e^{-\frac{1}{2(1-\rho^2)}\left[\left(\frac{x_1-\mu_{x_1}}{\sigma_{x_1}}\right)^2 - 2\rho\left(\frac{x_1-\mu_{x_1}}{\sigma_{x_1}}\right)\left(\frac{x_2-\mu_{x_2}}{\sigma_{x_2}}\right) + \left(\frac{x_2-\mu_{x_2}}{\sigma_{x_2}}\right)^2\right]}, \tag{15}$$

where $-\infty < x_i < \infty$ for $i$=1, 2 and $\rho = \frac{\sigma_{(x_1, x_2)}}{\sigma_{x_1}\sigma_{x_2}}$ is the correlation between $X_1$ and $X_2$. Now, define new RVs $Y_1 = e^{X_1}$ and $Y_2 = e^{X_2}$ such that $X_1 = \ln(Y_1)$ and $X_2 = \ln(Y_2)$. The Jacobian of this transformation is $\left(\frac{1}{Y_1}\frac{1}{Y_2}\right)$ and the joint lognormal PDF of $Y_1$ and $Y_2$ is defined using (15) as[2]:

$$f(y_1, y_2) =$$

$$\frac{1}{2\pi y_1 y_2 \sigma_{x_1}\sigma_{x_2}\sqrt{1-\rho^2}} e^{-\frac{1}{2(1-\rho^2)}\left[\left(\frac{\ln(y_1)-\mu_{x_1}}{\sigma_{x_1}}\right)^2 - 2\rho\left(\frac{\ln(y_1)-\mu_{x_1}}{\sigma_{x_1}}\right)\left(\frac{\ln(y_2)-\mu_{x_2}}{\sigma_{x_2}}\right) + \left(\frac{\ln(y_2)-\mu_{x_2}}{\sigma_{x_2}}\right)^2\right]}, \tag{16}$$

where $y_i \geq 0$ for $i$=1, 2. The means and variances for the lognormal RVs $Y_1$ and $Y_2$ are given by $E[Y_i] = e^{\mu_{x_i} + \frac{\sigma_{x_i}^2}{2}}$ and $V[Y_i] = \left(e^{2\mu_{x_i} + \sigma_{x_i}^2}\right)\left(e^{\sigma_{x_i}^2} - 1\right)$ for $i$=1, 2. The covariance is derived as (Law and Kelton [2000]):

$$\sigma_{(y_1, y_2)} = (e^{\sigma_{(x_1, x_2)}} - 1) * e^{\left(\mu_{x_1} + \mu_{x_2} + \frac{\sigma_{x_1}^2 + \sigma_{x_2}^2}{2}\right)}. \tag{17}$$

---

[2] The transformed joint PDF of $Y_1$ and $Y_2$, $f(y_1, y_2)$, is derived by substituting $X_1$ and $X_2$ with their expressions in terms of $Y_1$ and $Y_2$ in the original joint PDF, $f(x_1, x_2)$ (as shown in (15)), and multiplying by the absolute value of the Jacobian (Freund [1992]).

### D.4 Alternative Bivariate Form

As in the univariate case, natural logarithms are not required when defining bivariate lognormal RVs; any base is acceptable. In this research, we will use base 10 along with a constant factor. Assume $X_1$ and $X_2$ are joint normal RVs as defined in Section III.D.3. Let $\ddot{Y}_1 = 10^{X_1/10}$ and $\ddot{Y}_2 = 10^{X_2/10}$, which implies that $\ln(\ddot{Y}_1) = \theta X_1 \rightarrow \ddot{Y}_1 = e^{\theta X_1}$ and $\ln(\ddot{Y}_2) = \theta X_2 \rightarrow \ddot{Y}_2 = e^{\theta X_2}$. But, $\theta X_1 \sim N\left(\theta\mu_{x_1},\ \theta^2\sigma_{x_1}^2\right)$, and $\theta X_2 \sim N\left(\theta\mu_{x_2},\ \theta^2\sigma_{x_2}^2\right)$ with covariance and correlation as (Ross [2009]):

$$\text{Cov}(\theta X_1, \theta X_2) = \theta^2 \text{Cov}(X_1, X_2) = \theta^2 \sigma_{(x_1, x_2)}, \tag{18}$$

and,

$$\text{Corr}(\theta X_1, \theta X_2) = \frac{\theta^2 \sigma_{(x_1, x_2)}}{\theta^2 \sigma_{x_1} \sigma_{x_2}} = \frac{\sigma_{(x_1, x_2)}}{\sigma_{x_1} \sigma_{x_2}} = \rho, \tag{19}$$

as before. The correlation between the underlying normal RVs does not change form under the alternative representation. The joint lognormal PDF of $\ddot{Y}_1$ and $\ddot{Y}_2$ is therefore defined using (16) with the scaled means and variances as:

$$f(\ddot{y}_1, \ddot{y}_2) = \frac{\left(\frac{1}{\theta}\right)^2}{2\pi \ddot{y}_1 \ddot{y}_2 \sigma_{x_1} \sigma_{x_2}\sqrt{1-\rho^2}} *$$

$$e^{-\frac{1}{2(1-\rho^2)}\left[\left(\frac{\left(\frac{1}{\theta}\right)\ln(\ddot{y}_1)-\mu_{x_1}}{\sigma_{x_1}}\right)^2 - 2\rho\left(\frac{\left(\frac{1}{\theta}\right)\ln(\ddot{y}_1)-\mu_{x_1}}{\sigma_{x_1}}\right)\left(\frac{\left(\frac{1}{\theta}\right)\ln(\ddot{y}_2)-\mu_{x_2}}{\sigma_{x_2}}\right) + \left(\frac{\left(\frac{1}{\theta}\right)\ln(\ddot{y}_2)-\mu_{x_2}}{\sigma_{x_2}}\right)^2\right]}, \tag{20}$$

where $\ddot{y}_i \geq 0$ for $i$=1, 2. The means and variances for $\ddot{Y}_1$ and $\ddot{Y}_2$ are given by:

$$E[\ddot{Y}_i] = e^{\theta\mu_{x_i} + \frac{\theta^2\sigma_{x_i}^2}{2}}, \tag{21}$$

and,

$$V[\ddot{Y}_i] = \left(e^{2\theta\mu_{x_i} + \theta^2\sigma_{x_i}^2}\right) * \left(e^{\theta^2\sigma_{x_i}^2} - 1\right), \qquad \text{for } i = 1, 2. \tag{22}$$

The covariance is derived as (Law and Kelton [2000]):

$$\sigma_{(\ddot{y}_1, \ddot{y}_2)} = \left(e^{\theta^2 \sigma_{(x_1, x_2)}} - 1\right) * e^{\left(\theta(\mu_{x_1} + \mu_{x_2}) + \theta^2\left(\frac{\sigma_{x_1}^2 + \sigma_{x_2}^2}{2}\right)\right)}. \tag{23}$$

Lastly, if given $E[\ddot{Y}_i] = \mu_{\ddot{y}_i}$ and $V[\ddot{Y}_i] = \sigma^2_{\ddot{y}_i}$, $i$=1, 2, we can derive the underlying normal parameters by solving equations (21) and (22) for $\mu_{x_i}$ and $\sigma^2_{x_i}$, for $i$=1, 2. Doing so yields (Law and Kelton [2000]):

$$\mu_{x_i} = \left(\frac{1}{\theta}\right)\left(\ln(\mu_{\ddot{y}_i}) - \frac{1}{2}\ln\left[1 + \frac{\sigma^2_{\ddot{y}_i}}{\mu^2_{\ddot{y}_i}}\right]\right), \tag{24}$$

and,

$$\sigma^2_{x_i} = \left(\frac{1}{\theta}\right)^2 \ln\left[1 + \frac{\sigma^2_{\ddot{y}_i}}{\mu^2_{\ddot{y}_i}}\right], \tag{25}$$

for $i$=1, 2. Finally, if $Cov(\ddot{Y}_1, \ddot{Y}_2) = \sigma_{(\ddot{y}_1, \ddot{y}_2)}$ is known the covariance of the underlying normal RVs is given by (Law and Kelton [2000]):

$$\sigma_{(x_1, x_2)} = \left(\frac{1}{\theta}\right)^2 \ln\left[1 + \frac{\sigma_{(\ddot{y}_1, \ddot{y}_2)}}{|\mu_{\ddot{y}_1}\mu_{\ddot{y}_2}|}\right]. \tag{26}$$

### D.5  Moment Generating Function for the Lognormal Distribution

Let $X \sim N(\mu_x, \sigma_x{}^2)$ and $Y = e^X$ be the corresponding lognormal RV defined in standard form. Using (1) and (11) the MGF for Y is:

$$M_Y(t) = E_Y[e^{ty}] = \int_0^\infty e^{ty} * f(y)\, dy = \frac{1}{\sqrt{2\pi}\sigma_x} \int_0^\infty \frac{e^{\left[ty - \frac{1}{2}\left(\frac{\ln(y) - \mu_x}{\sigma_x}\right)^2\right]}}{y}\, dy. \tag{27}$$

The kernel of (27) has an indeterminate form of $\left(\frac{\infty}{\infty}\right)$ as $y \to \infty$ (for t > 0), since the power in the numerator can be expressed as $ty - (\frac{1}{2\sigma_x^2}[\ln(y)]^2 - \frac{\mu_x}{\sigma_x^2}\ln(y) + \frac{\mu_x^2}{2\sigma_x^2})$. Note that $ty - \frac{1}{2\sigma_x^2}[\ln(y)]^2$ increases without bound (for t > 0) as a function of $y$ since applying L'Hôpital's Rule twice to $\lim_{y \to \infty}\left(\frac{2\sigma_x^2 ty}{[\ln(y)]^2}\right)$ shows it to be infinite. This implies that $ty$ increases faster than $\frac{1}{2\sigma_x^2}[\ln(y)]^2$. To find which value the kernel in (27) approaches as $y$ increases, it suffices to consider the limit:

$$\lim_{y \to \infty}\left(\frac{e^{\left[ty - \frac{[\ln(y)]^2}{2\sigma_x^2}\right]}}{y}\right), \tag{28}$$

which, after one application of L'Hôpital's Rule (for t > 0), becomes:

$$\lim_{y \to \infty} \left( e^{\left[ ty - \frac{[\ln(y)]^2}{2\sigma_x^2} \right]} * \left[ t - \frac{1}{2\sigma_x^2} \frac{2\ln(y)}{y} \right] \right) = e^{\infty} * [t - 0] = \infty \,. \tag{29}$$

Since the kernel of (27) is $\geq 0$ and approaches $\infty$ as y increases (for $t > 0$), the area under it must also approach $\infty$ which implies that there does not exist an $\varepsilon > 0$ such the MGF is defined $\forall$ t $\in$ $(-\varepsilon, \varepsilon)$. This implies that the MGF of a lognormal RV does not exist. Casella and Berger (1990) refer to this as an "interesting property," namely that all moments for a lognormal RV exist and are finite, but, despite this fact, the MGF does not exist.

### D.6 Approximating the Lognormal Moment-Generating Function

Let X $\sim$ N($\mu_x$, $\sigma_x{}^2$) and $\ddot{Y} = 10^{X/10}$ be the alternative form lognormal RV as defined in Section III.D.2. Using the PDF from (14), the MGF for $\ddot{Y}$ is:

$$M_{\ddot{Y}}(t) = E_{\ddot{Y}}\left[ e^{t\ddot{y}} \right] = \int_0^\infty e^{t\ddot{y}} * f(\ddot{y}) \, d\ddot{y} = \int_0^\infty e^{t\ddot{y}} * \frac{1}{\theta \ddot{y} \sqrt{2\pi} \sigma_x} e^{-\frac{1}{2} \left( \frac{\left( \frac{1}{\theta} \right) \ln(\ddot{y}) - \mu_x}{\sigma_x} \right)^2} d\ddot{y} \,. \tag{30}$$

In the RHS integral from (30), we make the following U-substitution:

$$\text{Let } u = \frac{\left( \frac{1}{\theta} \right) \ln(\ddot{y}) - \mu_x}{\sqrt{2} \sigma_x} \tag{31}$$

$$\to \quad du = \frac{\left( \frac{1}{\theta} \right)}{\sqrt{2} \sigma_x} * \frac{1}{\ddot{y}} * d\ddot{y} \,, \tag{32}$$

and,

$$\ddot{y} = e^{\theta \left( u \sqrt{2} \sigma_x + \mu_x \right)} \,. \tag{33}$$

As $\ddot{y} \to 0$, u $\to -\infty$, and as $\ddot{y} \to \infty$, u $\to \infty$. Writing the integral from (30) in terms of $u$ yields the following expression for the MGF of $\ddot{Y}$:

$$M_{\ddot{Y}}(t) = \int_{-\infty}^\infty \frac{1}{\sqrt{\pi}} e^{t * e^{\theta \left( u\sqrt{2}\sigma_x + \mu_x \right)}} e^{-u^2} du = \int_{-\infty}^\infty g(u) * w(u) \, du \,. \tag{34}$$

Here, the weight function $w(u) = e^{-u^2}$ is of the form required by Gauss-Hermite quadrature, with appropriate integration limits, thus the MGF from (30) can be approximated (for t < 0) using the weights and roots provided in Table 1 with n=12. Namely,

$$M_{\ddot{Y}}(t) \;=\; \int_{-\infty}^{\infty} \frac{1}{\sqrt{\pi}} e^{t e^{\theta(u\sqrt{2}\sigma_x + \mu_x)}} e^{-u^2} du \;\approx\; \sum_{j=1}^{12} w_j * \frac{1}{\sqrt{\pi}} e^{t e^{\theta(t_j\sqrt{2}\sigma_x + \mu_x)}} . \qquad (35)$$

Recall that the MGF for a lognormal RV does not exist, something we showed in Section III.D.5, despite (35). The point is to approximate the sum of correlated lognormal RVs with a single lognormal RV which will be accomplished by equating their approximated MGFs for given t < 0, where it does exist. The justification for this is provided by Mitchell (1968) who showed that a single lognormal RV yields a better approximation to the sum than any other distribution examined, a property referred to as "permanence."

### D.7  *Approximating the Moment-Generating Function of a Sum*

Let $S = \alpha\ddot{Y}_1 + \beta\ddot{Y}_2$ be a weighted sum of two correlated lognormal RVs with bivariate PDF, $f(\ddot{y}_1, \ddot{y}_2)$, as shown in (20). The MGF for S is given by:

$$M_S(t) \;=\; E_S[e^{ts}] = E\!\left[e^{t(\alpha\ddot{y}_1 + \beta\ddot{y}_2)}\right] = \int_0^{\infty}\!\!\int_0^{\infty} e^{t(\alpha\ddot{y}_1 + \beta\ddot{y}_2)} f(\ddot{y}_1, \ddot{y}_2)\, d\ddot{y}_1 d\ddot{y}_2 . \qquad (36)$$

We will evaluate $M_S(t)$ in 2 steps. First, we make the following U-substitutions:

$$\text{Let } u_i = \left(\frac{1}{\theta}\right)\ln(\ddot{y}_i) \quad , for\ i = 1, 2 \qquad (37)$$

$$\rightarrow \quad du_i = \left(\frac{1}{\theta}\right)\frac{1}{\ddot{y}_i} d\ddot{y}_i \quad , for\ i = 1, 2\, , \qquad (38)$$

and,

$$\ddot{y}_i = e^{\theta u_i} . \qquad (39)$$

As $\ddot{y}_i \rightarrow 0$, $u_i \rightarrow -\infty$, and as $\ddot{y}_i \rightarrow \infty$, $u_i \rightarrow \infty$. The Jacobian matrix for this transformation has zeros on the off-diagonal, and terms $\theta e^{\theta u_i}$ on the i$^{\text{th}}$ diagonal, therefore the Jacobian determinant is equal to $\theta^2 e^{\theta u_1} e^{\theta u_2}$, and the right-side integral in (36) can be expressed using (20) as:

$$M_S(t) = \int\limits_{-\infty}^{\infty} \int\limits_{-\infty}^{\infty} e^{t(\alpha e^{\theta u_1} + \beta e^{\theta u_2})} \frac{1}{2\pi \sigma_{x_1} \sigma_{x_2} \sqrt{1-\rho^2}} *$$

$$e^{-\frac{1}{2(1-\rho^2)}\left[\left(\frac{u_1-\mu_{x_1}}{\sigma_{x_1}}\right)^2 - 2\rho\left(\frac{u_1-\mu_{x_1}}{\sigma_{x_1}}\right)\left(\frac{u_2-\mu_{x_2}}{\sigma_{x_2}}\right) + \left(\frac{u_2-\mu_{x_2}}{\sigma_{x_2}}\right)^2\right]} du_1 du_2 \qquad (40)$$

$$= \int\limits_{-\infty}^{\infty} \int\limits_{-\infty}^{\infty} e^{t(\alpha e^{\theta u_1} + \beta e^{\theta u_2})} f(u_1, u_2) \, du_1 du_2 \ . \qquad (41)$$

We have written $M_S(t)$ as the expected value of a function with respect to $U_1 \sim N(\mu_{x_1}, \sigma_{x_1}^2)$ and $U_2 \sim N(\mu_{x_2}, \sigma_{x_2}^2)$ with $\text{Corr}(U_1, U_2) = \rho$. In (41), $f(u_1, u_2)$ is a joint bivariate normal PDF with form as in (15). In matrix notation, this PDF can be expressed as (Guttman [1982]):

$$f(u_1, u_2) = \frac{1}{2\pi|\Sigma|^{1/2}} e^{-\frac{1}{2}(\vec{u}-\vec{\mu})' \Sigma^{-1} (\vec{u}-\vec{\mu})} \ , \quad for \ \vec{u} \in \Re^2 \ , \qquad (42)$$

where,

$$\vec{u} = \begin{pmatrix} u_1 \\ u_2 \end{pmatrix}, \ \ \vec{\mu} = \begin{pmatrix} \mu_{x_1} \\ \mu_{x_2} \end{pmatrix}, \ \ and, \ \ \Sigma = \begin{bmatrix} \sigma_{x_1}^2 & \sigma_{x_1 x_2} \\ \sigma_{x_1 x_2} & \sigma_{x_2}^2 \end{bmatrix} = \begin{bmatrix} \sigma_{x_1}^2 & \rho\sigma_{x_1}\sigma_{x_2} \\ \rho\sigma_{x_1}\sigma_{x_2} & \sigma_{x_2}^2 \end{bmatrix}. \qquad (43)$$

Here, $\Sigma$ is referred to as the variance-covariance matrix of $\vec{u}$ which we assume is symmetric and positive definite. Therefore, its inverse, $\Sigma^{-1}$, exists. Further, $\Sigma^{-1}$ will be symmetric and positive definite which implies there exist matrices $\mathbf{L}$ and $\mathbf{D}$ (where $\mathbf{L}$ is lower-triangular with 1's on the diagonal and $\mathbf{D}$ is diagonal with positive real pivots) such that (Meyer [2000]):

$$\Sigma^{-1} = \mathbf{LDL}' = \begin{bmatrix} 1 & 0 \\ \gamma & 1 \end{bmatrix} * \begin{bmatrix} d_{11} & 0 \\ 0 & d_{22} \end{bmatrix} * \begin{bmatrix} 1 & \gamma \\ 0 & 1 \end{bmatrix}. \qquad (44)$$

This is the LDU factorization of $\Sigma^{-1}$. We find $\mathbf{L}$ and $\mathbf{D}$ by equating terms as shown below (using the standard formula for a 2x2 matrix inverse) and solving for $d_{11}$, $d_{22}$, and $\gamma$:

$$\Sigma^{-1} = \frac{1}{\sigma_{x_1}^2 \sigma_{x_2}^2 (1-\rho^2)} \begin{bmatrix} \sigma_{x_2}^2 & -\rho\sigma_{x_1}\sigma_{x_2} \\ -\rho\sigma_{x_1}\sigma_{x_2} & \sigma_{x_1}^2 \end{bmatrix} = \mathbf{LDL}' = \begin{bmatrix} d_{11} & \gamma d_{11} \\ \gamma d_{11} & \gamma^2 d_{11} + d_{22} \end{bmatrix} \qquad (45)$$

$$\rightarrow \ \ d_{11} = \frac{1}{\sigma_{x_1}^2(1-\rho^2)}, \ \ \gamma = \frac{-\rho\sigma_{x_1}}{\sigma_{x_2}}, \ \ and, \ \ d_{22} = \frac{1}{\sigma_{x_2}^2}. \qquad (46)$$

The LDU factorization therefore yields:

$$\mathbf{\Sigma^{-1} = LDL'} = \begin{bmatrix} 1 & 0 \\ \dfrac{-\rho\sigma_{x_1}}{\sigma_{x_2}} & 1 \end{bmatrix} * \begin{bmatrix} \dfrac{1}{\sigma_{x_1}^2(1-\rho^2)} & 0 \\ 0 & \dfrac{1}{\sigma_{x_2}^2} \end{bmatrix} * \begin{bmatrix} 1 & \dfrac{-\rho\sigma_{x_1}}{\sigma_{x_2}} \\ 0 & 1 \end{bmatrix} \tag{47}$$

$$\mathbf{= LD^{0.5}D^{0.5}L'} = \begin{bmatrix} 1 & 0 \\ \dfrac{-\rho\sigma_{x_1}}{\sigma_{x_2}} & 1 \end{bmatrix} * \begin{bmatrix} \dfrac{1}{\sigma_{x_1}\sqrt{(1-\rho^2)}} & 0 \\ 0 & \dfrac{1}{\sigma_{x_2}} \end{bmatrix} * \begin{bmatrix} \dfrac{1}{\sigma_{x_1}\sqrt{(1-\rho^2)}} & 0 \\ 0 & \dfrac{1}{\sigma_{x_2}} \end{bmatrix} * \begin{bmatrix} 1 & \dfrac{-\rho\sigma_{x_1}}{\sigma_{x_2}} \\ 0 & 1 \end{bmatrix}. \tag{48}$$

The variance-covariance matrix $\mathbf{\Sigma}$ can be expressed similarly as:

$$\mathbf{\Sigma = (LDL')^{-1} = (L')^{-1}D^{-1}L^{-1}} = \begin{bmatrix} 1 & \dfrac{\rho\sigma_{x_1}}{\sigma_{x_2}} \\ 0 & 1 \end{bmatrix} * \begin{bmatrix} \sigma_{x_1}^2(1-\rho^2) & 0 \\ 0 & \sigma_{x_2}^2 \end{bmatrix} * \begin{bmatrix} 1 & 0 \\ \dfrac{\rho\sigma_{x_1}}{\sigma_{x_2}} & 1 \end{bmatrix} \tag{49}$$

$$= \begin{bmatrix} 1 & \dfrac{\rho\sigma_{x_1}}{\sigma_{x_2}} \\ 0 & 1 \end{bmatrix} * \begin{bmatrix} \sigma_{x_1}\sqrt{(1-\rho^2)} & 0 \\ 0 & \sigma_{x_2} \end{bmatrix} * \begin{bmatrix} \sigma_{x_1}\sqrt{(1-\rho^2)} & 0 \\ 0 & \sigma_{x_2} \end{bmatrix} * \begin{bmatrix} 1 & 0 \\ \dfrac{\rho\sigma_{x_1}}{\sigma_{x_2}} & 1 \end{bmatrix}. \tag{50}$$

The correlation between $U_1$ and $U_2$ originates from the correlation between $\breve{Y}_1$ and $\breve{Y}_2$ and prevents direct application of Gauss-Hermite quadrature to the MGF of $S = \alpha\breve{Y}_1 + \beta\breve{Y}_2$ as shown in (40). To address this, another transformation is made which decorrelates $U_1$ and $U_2$ (Mehta et al. [2007]) using the decomposition $\mathbf{\Sigma^{-1} = LDL'}$ shown in (47). A decorrelating transformation based on the LDU factorization of $\mathbf{\Sigma^{-1}}$ is:[3]

$$\text{Let } \vec{z} = \frac{1}{\sqrt{2}} \mathbf{D^{0.5}L'}(\vec{u} - \vec{\mu}), \text{ where } \vec{z} = \begin{pmatrix} z_1 \\ z_2 \end{pmatrix}. \tag{51}$$

The variance-covariance matrix for the random vector $\vec{z}$ is:

$$V(\vec{z}) = \frac{1}{2}\mathbf{D^{0.5}L'}V(\vec{u} - \vec{\mu})\mathbf{LD^{0.5}} = \frac{1}{2}\mathbf{D^{0.5}L'\,\Sigma\,LD^{0.5}} \tag{52}$$

$$= \frac{1}{2}\mathbf{D^{0.5}L'}[\mathbf{(L')^{-1}D^{-1}L^{-1}}]\,\mathbf{LD^{0.5}} \tag{53}$$

$$= \frac{1}{2}\mathbf{D^{0.5}\,D^{-1}D^{0.5}} = \frac{1}{2}\mathbf{I}. \tag{54}$$

Since the variance-covariance matrix, $V(\vec{z})$, of the transformed variables $Z_1$ and $Z_2$ is diagonal

---

[3] We have based this transformation on the LDU factorization of $\mathbf{\Sigma^{-1}}$. The matrix of this transformation is upper triangular, see (57). A similar decorrelating transformation can be obtained using the Cholesky decomposition of $\mathbf{\Sigma = LL'}$, namely, let $\vec{z} = \frac{1}{\sqrt{2}}\mathbf{L^{-1}}(\vec{u} - \vec{\mu})$. The matrix of this transformation, $\mathbf{L^{-1}}$, would then be lower triangular.

(i.e., (½)**I**), $Z_1$ and $Z_2$ are uncorrelated. Lastly, to compute the Jacobian of this transformation, we first express $\vec{u}$ in terms of $\vec{z}$ using (51), namely:

$$\vec{u} = \left(\sqrt{2}\right)(\mathbf{L}')^{-1}[\mathbf{D}^{0.5}]^{-1}\vec{z} + \vec{\mu} \tag{55}$$

$$\rightarrow \vec{u} = \left(\sqrt{2}\right)\begin{bmatrix} 1 & \dfrac{\rho\sigma_{x_1}}{\sigma_{x_2}} \\ 0 & 1 \end{bmatrix} * \begin{bmatrix} \sigma_{x_1}\sqrt{(1-\rho^2)} & 0 \\ 0 & \sigma_{x_2} \end{bmatrix} * \vec{z} + \vec{\mu}, \tag{56}$$

$$\rightarrow \vec{u} = \left(\sqrt{2}\right)\begin{bmatrix} \sigma_{x_1}\sqrt{(1-\rho^2)} & \rho\sigma_{x_1} \\ 0 & \sigma_{x_2} \end{bmatrix} * \vec{z} + \vec{\mu}, \tag{57}$$

so that,

$$u_1 = \sqrt{2}\left(\sigma_{x_1}\sqrt{(1-\rho^2)}\, z_1 + \rho\sigma_{x_1}z_2\right) + \mu_{x_1}, \tag{58}$$

and,

$$u_2 = \sqrt{2}\sigma_{x_2}z_2 + \mu_{x_2}. \tag{59}$$

The absolute value of the Jacobian determinant for this transformation is:

$$|\mathbf{J}| = \left| \mathrm{Det}\begin{bmatrix} \sqrt{2}\sigma_{x_1}\sqrt{(1-\rho^2)} & \sqrt{2}\rho\sigma_{x_1} \\ 0 & \sqrt{2}\sigma_{x_2} \end{bmatrix}\right| = 2\sigma_{x_1}\sigma_{x_2}\sqrt{(1-\rho^2)}. \tag{60}$$

Applying the decorrelating transformation we express the MGF of S from (40) in terms of $Z_1$ and $Z_2$ as:

$$M_S(t) = \int_{-\infty}^{\infty} \int_{-\infty}^{\infty} e^{t*\alpha*e^{\theta\left[\sqrt{2}\left(\sigma_{x_1}\sqrt{(1-\rho^2)}\, z_1 + \rho\sigma_{x_1}z_2\right)+\mu_{x_1}\right]}} e^{t*\beta*e^{\theta\left[\sqrt{2}\sigma_{x_2}z_2+\mu_{x_2}\right]}} f(z_1,z_2)dz_1dz_2, \tag{61}$$

where,

$$f(z_1,z_2) = \frac{\sigma_{x_1}\sigma_{x_2}\sqrt{(1-\rho^2)}}{\pi|(\mathbf{L}')^{-1}\mathbf{D}^{-1}\mathbf{L}^{-1}|^{1/2}}e^{-\vec{z}'\,\vec{z}} = \frac{\sigma_{x_1}\sigma_{x_2}\sqrt{(1-\rho^2)}}{\pi|\mathbf{D}^{-1}|^{1/2}}e^{-\vec{z}'\,\vec{z}} \tag{62}$$

$$= \frac{1}{\pi}e^{-\vec{z}'\,\vec{z}} = \frac{1}{\pi}e^{-z_1^2}e^{-z_2^2}, \quad for \ \vec{z} \in \Re^2. \tag{63}$$

The MGF of S $= \alpha\ddot{Y}_1 + \beta\ddot{Y}_2$ from (40) can therefore be expressed in terms of $Z_1$ and $Z_2$ as:

$$M_S(t) = \int_{-\infty}^{\infty} \int_{-\infty}^{\infty} \frac{1}{\pi}e^{t*\alpha*e^{\theta\left[\sqrt{2}\left(\sigma_{x_1}\sqrt{(1-\rho^2)}+\rho\sigma_{x_1}z_2\right)+\mu_{x_1}\right]}} e^{t*\beta*e^{\theta\left[\sqrt{2}\sigma_{x_2}z_2+\mu_{x_2}\right]}} e^{-z_1^2}e^{-z_2^2}\, dz_1dz_2 \tag{64}$$

$$\rightarrow \quad M_S(t) = \int\limits_{-\infty}^{\infty} \int\limits_{-\infty}^{\infty} \frac{1}{\pi} h(z_1, z_2) \, e^{-z_1^2} e^{-z_2^2} \, dz_1 dz_2 \quad , \tag{65}$$

where,

$$h(z_1, z_2) = e^{t*\alpha*e^{\theta\left[\sqrt{2}\left(\sigma_{x_1}\sqrt{(1-\rho^2)}\, z_1 + \rho\sigma_{x_1} z_2\right) + \mu_{x_1}\right]}} e^{t*\beta*e^{\left[\sqrt{2}\sigma_{x_2} z_2 + \mu_{x_2}\right]}} . \tag{66}$$

The representation of $M_S(t)$ in (65) is of the form required for Gauss-Hermite quadrature using the weights and roots from Table 1. The MGF of $S = \alpha \ddot{Y}_1 + \beta \ddot{Y}_2$ can therefore be approximated in 2 steps. In Step 1, we apply the quadrature rules to $Z_1$ and replace the integral by a sum, then we repeat the process for $Z_2$ in Step 2.

<u>Step 1:</u> Apply Gauss-Hermite Quadrature Rules (n=12) to $Z_1$

$$\rightarrow \quad M_S(t) \cong \frac{1}{\pi} \int\limits_{-\infty}^{\infty} \left( \sum_{j=1}^{12} w_j * \left[ h(t_j, z_2) \right] \right) e^{-z_2^2} \, dz_2 \tag{67}$$

<u>Step 2:</u> Apply Gauss-Hermite Quadrature Rules (n=12) to $Z_2$

$$\rightarrow \quad M_S(t) \cong \frac{1}{\pi} \sum_{i=1}^{12} w_i * \left( \sum_{j=1}^{12} w_j * \left[ h(t_j, t_i) \right] \right) \tag{68a}$$

$$\rightarrow \quad M_S(t) \cong \frac{1}{\pi} \sum_{i=1}^{12} \sum_{j=1}^{12} w_i * w_j * h(t_j, t_i) \tag{68b}$$

The function $h(z_1, z_2)$ is given in (66) and the final step is to equate $M_S(t)$ from (68b) to the approximated MGF for a univariate lognormal RV $M_{\ddot{Y}}(t)$, shown in the RHS of (35), and solve for the 2 unknowns $\mu_x$ and $\sigma_x$. It is assumed that $\mu_{x_1}$, $\mu_{x_2}$, $\sigma_{x_1}$, $\sigma_{x_2}$, and $\rho$ are all known constants. Since 2 equations are needed to obtain a solution for the two unknowns, we generate these equations using different real values for t < 0 (Mehta et al. [2007]).

## IV.  Approximation of Lognormal Sum with Lognormal RV

In this research, we have correlated lognormal RVs $\ddot{Y}_1$ and $\ddot{Y}_2$ with $\text{Cov}(\ddot{Y}_1, \ddot{Y}_2) = \sigma_{(\ddot{y}_1, \ddot{y}_2)}$.  The means and variances are known and given by $\text{E}[\ddot{Y}_1] = \mu_{\ddot{y}_1}$, $\text{E}[\ddot{Y}_2] = \mu_{\ddot{y}_2}$, $\text{V}[\ddot{Y}_1] = \sigma_{\ddot{y}_1}^2$, and $\text{V}[\ddot{Y}_2] = \sigma_{\ddot{y}_2}^2$.  Both $\ddot{Y}_1$ and $\ddot{Y}_2$ are defined in non-standard form, so there exist normally distributed RVs $X_1 \sim N(\mu_{x_1}, \sigma_{x_1}^2)$ and $X_2 \sim N(\mu_{x_2}, \sigma_{x_2}^2)$ such that $\ddot{Y}_1 = 10^{X_1/10}$ and $\ddot{Y}_2 = 10^{X_2/10}$ where $\text{Corr}(X_1, X_2) = \rho$.  We will not be provided the means and variances of these underlying normal RVs, but they will be calculated using the expressions in (24), (25), and (26).  Lastly, we will be given constants $\alpha$ and $\beta$ and have interest in approximating the probability distribution of $S = \alpha\ddot{Y}_1 + \beta\ddot{Y}_2$.  Based on Mitchell (1968), using a univariate lognormal RV to approximate the distribution of $S$ is desirable.  We will first calculate the parameters for the underlying normal RVs and then set $M_S(t)$ from (68b) equal to $M_{\ddot{Y}}(t)$ from (35) and solve for $\mu_x$ and $\sigma_x$.  These are the mean and standard deviation, respectively, from the underlying normal distribution that forms the base distribution of our lognormal approximation.  The final step is to calculate the corresponding mean and variance for the approximating univariate lognormal distribution using the expressions in (12) and (13).

### A.  An Example

Let $\ddot{Y}_1$ and $\ddot{Y}_2$ be (non-standard form) lognormal RVs with $\mu_{\ddot{y}_1} = 1.0$, $\mu_{\ddot{y}_2} = 2.0$, $\sigma_{\ddot{y}_1}^2 = 3.0$, $\sigma_{\ddot{y}_2}^2 = 4.0$, and $\sigma_{(\ddot{y}_1, \ddot{y}_2)} = 1.73$ so that $\text{Corr}(\ddot{Y}_1, \ddot{Y}_2) = \sigma_{(\ddot{y}_1, \ddot{y}_2)} / (\sigma_{\ddot{y}_1}\sigma_{\ddot{y}_2}) = 0.5$.  Further, define constants $\alpha = 1.5$ and $\beta = 2.5$ with interest in approximating the probability distribution of $S = \alpha\ddot{Y}_1 + \beta\ddot{Y}_2 = (1.5)\ddot{Y}_1 + (2.5)\ddot{Y}_2$.  Using (24), (25), and (26) from Section III.D.4, the parameters of the underlying normal distributions are given by:

$\rightarrow$ From Equation (24):

$$\mu_{x_1} = \left(\frac{1}{\theta}\right)\left(\ln(1.0) - \frac{1}{2}\ln\left[1 + \frac{3.0}{(1.0)^2}\right]\right) = -3.0103 \tag{69}$$

$$\mu_{x_2} = \left(\frac{1}{\theta}\right)\left(\ln(2.0) - \frac{1}{2}\ln\left[1 + \frac{4.0}{(2.0)^2}\right]\right) = 1.5051 \tag{70}$$

$\rightarrow$ From Equation (25):

$$\sigma_{x_1}^2 = \left(\frac{1}{\theta}\right)^2 \ln\left[1 + \frac{3.0}{(1.0)^2}\right] = 26.1471 \tag{71}$$

$$\sigma_{x_2}^2 = \left(\frac{1}{\theta}\right)^2 \ln\left[1 + \frac{4.0}{(2.0)^2}\right] = 13.0736 \tag{72}$$

$\rightarrow$ From Equation (26):

$$\sigma_{(x_1, x_2)} = \left(\frac{1}{\theta}\right)^2 \ln\left[1 + \frac{1.73}{|1.0 * 2.0|}\right] = 11.7554 \tag{73}$$

Therefore, Corr($X_1$, $X_2$) = $\rho = \frac{11.7554}{\sqrt{26.1471 * 13.0736}} = 0.635811$, and $1 - \rho^2 = 0.595744$. Using these

quantities the function $h(z_1, z_2)$ from (66) becomes:

$$\tag{74}$$

$$h(z_1, z_2) = e^{(1.5)te^{\theta\left[\sqrt{2}\left((5.1134)\sqrt{0.595744}\,z_1 + (0.635811)(5.1134)z_2\right) + (-3.0103)\right]}} e^{(2.5)te^{\theta\left[\sqrt{2}(3.6157)z_2 + (1.5051)\right]}}$$

When CDF values of the sum S are desired, Mehta et al. (2007) find good results using constants $t_1 = -1.0$ and $t_2 = -0.2$ for t to generate the needed equations.[4] Using these values, and setting $M_S(t)$ from (68) equal to $M_{\hat{Y}}(t)$ from (35), we solve the following two non-linear equations for a single $\mu_x$ and $\sigma_x$, which are the only two unknown quantities in equations (75) and (76). The quadrature weights and roots, $w_i$, $w_j$, $t_i$, and $t_j$ are provided in Table 1.

**Equation #1:**

$$\frac{1}{\pi}\sum_{i=1}^{12}\sum_{j=1}^{12}\left[w_i * w_j * e^{(-1.5)*e^{\theta\left[\sqrt{2}\left((5.1134)\sqrt{0.595744}t_j + (0.635811)(5.1134)t_i\right) + (-3.0103)\right]}} * \right.$$
$$\left. e^{(-2.5)*e^{\theta\left[\sqrt{2}(3.6157)t_i + (1.5051)\right]}}\right] = \frac{1}{\sqrt{\pi}}\sum_{j=1}^{12}w_j * e^{(-1.0)*e^{\theta\left(\sqrt{2}\sigma_x t_j + \mu_x\right)}} \tag{75}$$

**Equation #2:**

$$\frac{1}{\pi}\sum_{i=1}^{12}\sum_{j=1}^{12}\left[w_i * w_j * e^{(-0.3)*e^{\theta\left[\sqrt{2}\left((5.1134)\sqrt{0.595744}t_j + (0.635811)(5.1134)t_i\right) + (-3.0103)\right]}} * \right.$$
$$\left. e^{(-0.5)*e^{\theta\left[\sqrt{2}(3.6157)t_i + (1.5051)\right]}}\right] = \frac{1}{\sqrt{\pi}}\sum_{j=1}^{12}w_j * e^{(-0.2)*e^{\theta\left(\sqrt{2}\sigma_x t_j + \mu_x\right)}} \tag{76}$$

---

[4] These values are for the quantity t as defined from the MGF which is different from the $t_i$ and $t_j$ values used to denote the roots for Gauss-Hermite quadrature which are provided in Table 1.

## B. Implementation Details

Equations 1 and 2 from (75) and (76) must be simultaneously solved for $\mu_x$ and $\sigma_x$ but note that the left-hand sides of both equations are known constants. Each is a $12^2 = 144$ term sum with no unknown quantities involved. The left-hand sides can thus be subtracted and the equations expressed as simultaneous non-linear equations equal to zero. Further, the equations were generated using $t \in \{-1.0, -0.2\}$ as suggested by Mehta et al. (2007). In general terms let $t \in \{\tau_1, \tau_2\}$, and let $C_1$ and $C_2$ be the LHS of (75) and (76), respectively. It follows that these equations can be expressed as:

**Equation #1:**

$$\frac{1}{\sqrt{\pi}} \sum_{j=1}^{12} w_j * e^{(\tau_1)e^{\theta\left(\sqrt{2}\sigma_x t_j + \mu_x\right)}} - C_1 = 0 \qquad (77)$$

**Equation #2:**

$$\frac{1}{\sqrt{\pi}} \sum_{j=1}^{12} w_j * e^{(\tau_2)e^{\theta\left(\sqrt{2}\sigma_x t_j + \mu_x\right)}} - C_2 = 0 \qquad (78)$$

The constants $C_1$ and $C_2$ in (77) and (78) are specific to the problem being addressed, and $\{\tau_1, \tau_2\}$ may also be application specific. If we denote the LHS of (77) and (78) by $M_{\check{Y}}^{(1)}(\mu_x, \sigma_x)$ and $M_{\check{Y}}^{(2)}(\mu_x, \sigma_x)$, respectively, then we seek to solve the following non-linear system for $\mu_x$ and $\sigma_x$:

$$\vec{\mathbf{m}} = \begin{pmatrix} M_{\check{Y}}^{(1)}(\mu_x, \sigma_x) \\ M_{\check{Y}}^{(2)}(\mu_x, \sigma_x) \end{pmatrix} = \begin{pmatrix} 0 \\ 0 \end{pmatrix}. \qquad (79)$$

Newton's method is often used in optimization problems to solve a similar set of non-linear equations, namely that of the gradient vector being equal to zero. It applies here as well and operates by approximating the vector $\vec{\mathbf{m}}$ by its $1^{st}$ order Taylor expansion around some given initial starting point $(\mu_x^0, \sigma_x^0)$, namely:

$$\vec{\mathbf{m}} \cong \vec{\mathbf{m}}_0 = \begin{pmatrix} M_{\check{Y}}^{(1)}(\mu_x^0, \sigma_x^0) \\ M_{\check{Y}}^{(2)}(\mu_x^0, \sigma_x^0) \end{pmatrix} + \begin{bmatrix} \frac{\partial}{\partial \mu_x}\left[M_{\check{Y}}^{(1)}\right]_{(\mu_x^0, \sigma_x^0)} & \frac{\partial}{\partial \sigma_x}\left[M_{\check{Y}}^{(1)}\right]_{(\mu_x^0, \sigma_x^0)} \\ \frac{\partial}{\partial \mu_x}\left[M_{\check{Y}}^{(2)}\right]_{(\mu_x^0, \sigma_x^0)} & \frac{\partial}{\partial \sigma_x}\left[M_{\check{Y}}^{(2)}\right]_{(\mu_x^0, \sigma_x^0)} \end{bmatrix} * \begin{pmatrix} \mu_x - \mu_x^0 \\ \sigma_x - \sigma_x^0 \end{pmatrix}. \qquad (80)$$

The matrix shown in (80) consists of the corresponding $1^{st}$ order partial derivatives of both functions evaluated at the initial starting point, and it is therefore known once the derivatives have been calculated. By setting this $1^{st}$ order Taylor expansion of $\vec{\mathbf{m}}$ at $(\mu_x^0, \sigma_x^0)$ (i.e., $\vec{\mathbf{m}}_0$), equal to zero and solving for $(\mu_x, \sigma_x)$, we arrive at:

$$\vec{\mathbf{m}}_0 = \begin{pmatrix} 0 \\ 0 \end{pmatrix} \tag{81}$$

$$\leftrightarrow \begin{bmatrix} \dfrac{\partial}{\partial \mu_x}\left[M_{\mathring{Y}}^{(1)}\right]_{(\mu_x^0, \sigma_x^0)} & \dfrac{\partial}{\partial \sigma_x}\left[M_{\mathring{Y}}^{(1)}\right]_{(\mu_x^0, \sigma_x^0)} \\ \dfrac{\partial}{\partial \mu_x}\left[M_{\mathring{Y}}^{(2)}\right]_{(\mu_x^0, \sigma_x^0)} & \dfrac{\partial}{\partial \sigma_x}\left[M_{\mathring{Y}}^{(2)}\right]_{(\mu_x^0, \sigma_x^0)} \end{bmatrix} * \begin{pmatrix} \mu_x - \mu_x^0 \\ \sigma_x - \sigma_x^0 \end{pmatrix} = -\begin{pmatrix} M_{\mathring{Y}}^{(1)}(\mu_x^0, \sigma_x^0) \\ M_{\mathring{Y}}^{(2)}(\mu_x^0, \sigma_x^0) \end{pmatrix} \tag{82}$$

which is a simple linear system of 2 equations and 2 unknowns that we solve for the vector:

$$\begin{pmatrix} \mu_x - \mu_x^0 \\ \sigma_x - \sigma_x^0 \end{pmatrix}. \tag{83}$$

The solution immediately yields values for $(\mu_x, \sigma_x)$ that we label $(\mu_x^1, \sigma_x^1)$ and then the process is repeated using $(\mu_x^1, \sigma_x^1)$ as the new starting point which yields $(\mu_x^2, \sigma_x^2)$, and so on. We stop at iteration $i$ when an objective criteria is met, such as when the functions $M_{\mathring{Y}}^{(1)}(\mu_x^i, \sigma_x^i)$ and $M_{\mathring{Y}}^{(2)}(\mu_x^i, \sigma_x^i)$ both become smaller than some predetermined threshold level $\varepsilon$. The only remaining step for implementing Newton's method is to derive the elements of the 2x2 matrix of partial derivatives shown in (80). Using the chain rule along with the fact that the derivative of a sum equals the sum of the derivatives, these quantities are given by:

$$\frac{\partial}{\partial \mu_x}\left[M_{\mathring{Y}}^{(i)}(\mu_x, \sigma_x)\right] = \frac{1}{\sqrt{\pi}} \sum_{j=1}^{12} w_j * \frac{\partial}{\partial \mu_x}\left[e^{(\tau_i)e^{\theta\left(\sqrt{2}\sigma_x t_j + \mu_x\right)}}\right] \tag{84a}$$

$$= \frac{\theta * \tau_i}{\sqrt{\pi}} \sum_{j=1}^{12} w_j * e^{(\tau_i)e^{\theta\left(\sqrt{2}\sigma_x t_j + \mu_x\right)}} * e^{\theta\left(\sqrt{2}\sigma_x t_j + \mu_x\right)} \tag{84b}$$

$$\text{for } i = 1, 2,$$

and,

$$\frac{\partial}{\partial \sigma_x}\left[M_{\ddot{Y}}^{(i)}(\mu_x, \sigma_x)\right] = \frac{1}{\sqrt{\pi}}\sum_{j=1}^{12} w_j * \frac{\partial}{\partial \sigma_x}\left[e^{(\tau_i)e^{\theta\left(\sqrt{2}\sigma_x t_j + \mu_x\right)}}\right] \tag{85a}$$

$$= \frac{\theta * \tau_i * \sqrt{2}}{\sqrt{\pi}}\sum_{j=1}^{12} w_j * t_j * e^{(\tau_i)e^{\theta\left(\sqrt{2}\sigma_x t_j + \mu_x\right)}} * e^{\theta\left(\sqrt{2}\sigma_x t_j + \mu_x\right)} \tag{85b}$$

$$\text{for } i = 1, 2.$$

The 1$^{st}$ order partial derivatives above are evaluated at the starting point for each iteration, thus they are completely known and constitute the 2x2 matrix from (80). All quantities in the system of 2 linear equations from (82) are known except $(\mu_x, \sigma_x)$ and we will solve for these using the technique described. Solving a linear system of 2 equations with 2 unknowns is trivial. Once found, the mean and variance of the approximating univariate lognormal RV for $S = \alpha\ddot{Y}_1 + \beta\ddot{Y}_2$, namely E[S] and V[S], are derived using (12) and (13).

### B.1 Initial Values for Newton's Method

Newton's method requires the selection of initial values, namely $(\mu_x^0, \sigma_x^0)$, and choosing values that are closer to the actual solution can reduce the number of iterations needed to achieve convergence. Since we are interested in the sum $S = \alpha\ddot{Y}_1 + \beta\ddot{Y}_2$, an obvious choice, motivated by the F-W approximation, is to use the values that correspond to E[S] and V[S], both of which are known. That is,

$$E[S] = E[\alpha\ddot{Y}_1 + \beta\ddot{Y}_2] = \alpha\mu_{\ddot{y}_1} + \beta\mu_{\ddot{y}_2} \tag{86}$$

$$V[S] = V\left[(\alpha \ \ \beta) * \binom{\ddot{Y}_1}{\ddot{Y}_2}\right] = (\alpha \ \ \beta)V\binom{\ddot{Y}_1}{\ddot{Y}_2}\binom{\alpha}{\beta} = (\alpha \ \ \beta)\begin{bmatrix} \sigma_{\ddot{y}_1}^2 & \sigma_{(\ddot{y}_1, \ddot{y}_2)} \\ \sigma_{(\ddot{y}_1, \ddot{y}_2)} & \sigma_{\ddot{y}_2}^2 \end{bmatrix}\binom{\alpha}{\beta}. \tag{87}$$

The initial values are then derived using (24) and (25) as:

$$\mu_x^0 = \left(\frac{1}{\theta}\right)\left(\ln(E[S]) - \frac{1}{2}\ln\left[1 + \frac{V[S]}{(E[S])^2}\right]\right), \tag{88}$$

$$\sigma_x^0 = \sqrt{\left(\frac{1}{\theta}\right)^2 \ln\left[1 + \frac{V[S]}{(E[S])^2}\right]}. \tag{89}$$

## V. An Application to Finance

Let R be the total annual return on an arbitrary investment portfolio.  For prior years, we can calculate the value of R as:

$$R = \frac{\text{End Balance} - \text{Start Balance}}{\text{Start Balance}} \ . \tag{90}$$

From (90), the portfolio's ending balance can be derived using R as:

$$\text{End Balance} = (\text{Start Balance}) * (1 + R) \ .$$

Future unobserved values of R will be taken as RVs following some probability distribution. The quantity R consists of an inflation component, I, and a real return component, r, and it can be decomposed as follows:

$$(1 + R) = (1 + I) * (1 + r) \ . \tag{91}$$

Here, both I and r are RVs but the inflation rate can be difficult to model because it possesses a deterministic component.  The inflation rate is heavily influenced by central banks via monetary policy, which can make treating it as a pure RV problematic.  Further, central banks often have a target inflation rate and the process of keeping it on target can add serial correlation to the observations.  To remove it from the model, we divide both sides of (91) by (1 + I), which leaves:

$$(1 + r) = \frac{(1 + R)}{(1 + I)} \ . \tag{92}$$

The quantity r is referred to as the real (or inflation-adjusted) annual return on the investment portfolio and it can be positive or negative.  The value $(1 + r)$ is the compounding real annual return and it must be $\geq 0$ since our investment portfolio can lose all of its value in a single year, but it cannot have a negative balance.  If we treat r as a normally distributed RV, then $(1 + r)$ is also normally distributed with non-zero probability of taking a negative value.  For this reason, it may be preferable to assume that $(1 + r)$ is lognormally distributed.  The lognormal distribution can also be justified by viewing the annual compounding returns as the product of daily compounding returns.  It was noted earlier that, via the CLT, the product of more than 30 *iid* RVs is approximately lognormal.   To find the best fitting lognormal distribution for $(1 + r)$, we examine the historical record.  For example, if we invest our portfolio in an S&P 500 Index Fund or in 10-Year Treasury Bonds, then the historical record will reveal the annual total returns $R_t$ for each investment, along with the inflation rate $I_t$, and it is a simple matter to construct $r_t$ for time points t=1, 2, …, N, as:

$$r_t = \frac{(1 + R_t)}{(1 + I_t)} - 1 \ . \tag{93}$$

The values $(1 + r_t)$ can be fit to a lognormal distribution using a hypothesis test such as the Anderson-Darling (A/D) test. The null hypothesis is that the returns originate from a given lognormal distribution and a p-value is generated. For example, using historical data on the S&P 500 Index (stocks) and 10-Year Treasury Bonds (bonds), along with the corresponding inflation rates from 1928 – 2013, the real compounding returns $(1 + r_s)$ and $(1 + r_b)$, for stocks and bonds, respectively, are best fit by the following LogNormal($\mu, \sigma$) distributions[5]:

$$(1 + r_s) \sim \text{LogNormal}(1.0837, 0.2153) \quad (\text{A/D} \ \ p - \text{value} = 0.000) \tag{94}$$

$$(1 + r_b) \sim \text{LogNormal}(1.0214, 0.0825) \quad (\text{A/D} \ \ p - \text{value} = 0.559) \tag{95}$$

As shown, S&P 500 Index real compounding returns, $(1 + r_s)$, have a p-value that leads to rejection of the null hypothesis that they originate from a lognormal distribution, perhaps suggesting that the daily returns are not *iid*. The corresponding hypothesis for 10-Year Treasury Bond returns cannot be rejected at any reasonable significance level. Note that similar null hypotheses for both $(1 + r_s)$ and $(1 + r_b)$ with respect to the normal distribution cannot be rejected at any reasonable significance level. Regardless, we will accept this disparity for the benefit of using RVs that have a domain which is consistent with the practical application. Finally, the sample correlation and covariance between these real compounding returns, at a given time point, is measured as:

$$\text{Corr}[(1 + r_s), (1 + r_b)] = 0.04387 \tag{96}$$

$$\text{Cov}[(1 + r_s), (1 + r_b)] = 0.00078 \tag{97}$$

Let $R_s$ and $R_b$ be total annual returns for the stock and bond investments detailed above, respectively. A diversified portfolio would invest the proportion α in stocks and (1-α) in bonds. The total annual return on this portfolio is $αR_s$ + (1-α)$R_b$ and the corresponding compounding return is $(1 + \alpha R_s + (1-\alpha)R_b) = \alpha(1 + R_s) + (1-\alpha)(1 + R_b)$.  By decomposing each total return into its inflation and real component the compounding return can be written as $\alpha(1 + r_s)(1 + I) + (1-\alpha)(1 + r_b)(1 + I)$.  To obtain the compounding real return, we divide by $(1 + I)$ which yields $\alpha(1 + r_s) + (1-\alpha)(1 + r_b) = (1 + \alpha r_s + (1-\alpha)r_b)$.  This is a weighted sum of correlated lognormal RVs, see (94) and (95).  Since $\alpha$ is the proportion invested in stocks, it is often referred to as the equity ratio.  The CDF of real compounding returns on a diversified stock and bond portfolio can thus be approximated using the techniques presented here.

[5] S&P 500 Index & 10-Year Treasury Bond total returns were retrieved from NYU Professor Aswath Damodaran's financial database which can be accessed at: http://pages.stern.nyu.edu/~adamodar/. The corresponding inflation rates were taken as the CPI-U and retrieved from the U.S. Federal Reserve Bank of Minneapolis website which can be accessed at: http://www.minneapolisfed.org/community_education/teacher/calc/hist1913.cfm.

Consider diversified portfolios consisting of equity ratios $\alpha \in \{0.25, 0.50, 0.75\}$.  Probabilities for the compounding return $S = (1 + \alpha r_s + (1-\alpha)r_b)$ will be derived using the MGF technique presented and compared with simulated probabilities and probabilities derived from the moment-matching (M-M) lognormal distribution[6].  We will examine probabilities from both the head and tail along with those near the mean.  The method presented here will use $t \in \{(-1.0, -0.2), (-0.001, -0.005)\}$ as proposed by Mehta et al. (2007), who note that some t-sets work better in the head portion, and others in the tail of the distribution of S.  The results of this analysis are shown below in Table 2.

**Table 2**
**Comparison of CDF Probabilities using Various Methods**[a]

| Method | Equity Ratio ($\alpha$) | CDF Probabilities $P(S \leq s)$ | | | | | | | | |
|---|---|---|---|---|---|---|---|---|---|---|
| | | **0.01** | **0.05** | **0.10** | **0.30** | **0.50** | **0.80** | **0.90** | **0.95** | **0.99** |
| Simulation[b] | 0.25 | 0.8589 | 0.9063 | 0.9327 | 0.9906 | 1.0322 | 1.1061 | 1.1463 | 1.1811 | 1.2498 |
| | 0.50 | 0.8202 | 0.8778 | 0.9108 | 0.9861 | 1.0434 | 1.1463 | 1.2063 | 1.2591 | 1.3683 |
| | 0.75 | 0.7536 | 0.8280 | 0.8721 | 0.9735 | 1.0530 | 1.1982 | 1.2840 | 1.3605 | 1.5198 |
| M-M | 0.25 | 0.8568 | 0.9052 | 0.9321 | 0.9908 | 1.0336 | 1.1062 | 1.1462 | 1.1802 | 1.2469 |
| | 0.50 | 0.8084 | 0.8718 | 0.9077 | 0.9871 | 1.0461 | 1.1483 | 1.2057 | 1.2552 | 1.3536 |
| | 0.75 | 0.7407 | 0.8218 | 0.8685 | 0.9747 | 1.0558 | 1.2002 | 1.2834 | 1.3565 | 1.5049 |
| MGF(1)[c] | 0.25 | 0.8569 | 0.9053 | 0.9322 | 0.9908 | 1.0336 | 1.1062 | 1.1461 | 1.1801 | 1.2468 |
| | 0.50 | 0.8093 | 0.8725 | 0.9082 | 0.9873 | 1.0462 | 1.1480 | 1.2051 | 1.2544 | 1.3524 |
| | 0.75 | 0.7418 | 0.8226 | 0.8693 | 0.9751 | 1.0559 | 1.1997 | 1.2826 | 1.3553 | 1.5029 |
| MGF(2)[c] | 0.25 | 0.8568 | 0.9052 | 0.9321 | 0.9908 | 1.0336 | 1.1062 | 1.1462 | 1.1802 | 1.2469 |
| | 0.50 | 0.8084 | 0.8718 | 0.9077 | 0.9871 | 1.0461 | 1.1483 | 1.2057 | 1.2552 | 1.3536 |
| | 0.75 | 0.7407 | 0.8218 | 0.8685 | 0.9747 | 1.0558 | 1.2002 | 1.2834 | 1.3565 | 1.5049 |

[a] Probabilities are for $S = \alpha \tilde{Y}_1 + (1-\alpha)\tilde{Y}_2$ where $\tilde{Y}_1 \sim LogNormal(1.0837, 0.2153)$, $\tilde{Y}_2 \sim LogNormal(1.0214, 0.0825)$ and $Cov(\tilde{Y}_1, \tilde{Y}_2) = 0.00078$.  The cell values represent the lognormal domain values, s, that yield $P(S \leq s)$.
[b] Simulations were run in C++ using a sample size of $N = 200,000,000$.
[c] MGF(1) uses $t \in \{-1.0, -0.2\}$ and MGF(2) uses $t \in \{-0.001, -0.005\}$.

---

[6] The moment-matching lognormal distribution would be the one with mean and variance equal to E[S] and V[S], respectively, and derived in (86) and (87).  It is motivated by the F-W approach for sums of independent RVs.

The CDF values from Table 2 generated via simulation can be viewed as the best representation of the true probabilities. The first item of note from Table 2 is that the CDF probabilities using the M-M lognormal approximation and using the MGF technique presented here with t ∈ {-0.001, -0.005} are identical. When values of t near zero are used, the equations from (77) and (78) are instantly satisfied without iterating and the procedure converges to the initial values. To see this, note that when t ∈ {0.0, 0.0}, $h(z_1, z_2)$ from (66) equals 1, and $C_1$, $C_2$ become:

$$C_1 = C_2 = \frac{1}{\pi} \sum_{i=1}^{12} \sum_{j=1}^{12} [w_i * w_j] . \tag{98}$$

Further, equations (77) and (78) are identical and both reduce to:

$$\frac{1}{\sqrt{\pi}} \sum_{j=1}^{12} w_j - \frac{1}{\pi} \sum_{i=1}^{12} \sum_{j=1}^{12} [w_i * w_j] = 0 . \tag{99}$$

But since,

$$\sum_{j=1}^{12} w_j \approx \sqrt{\pi}, \quad \text{and,} \quad \sum_{i=1}^{12} \sum_{j=1}^{12} [w_i * w_j] \approx \pi , \tag{100}$$

the equation in (99) is automatically satisfied, thus converges at the initial values. For this reason, using two values of t near zero is not recommended, as any initial values satisfy the equations and the procedure converges instantly to these values. It is straightforward to prove the results from (100). Note that the area under a standard normal RV equals 1 since it is a valid PDF. Let z ~ N(0,1), then:

$$\int_{-\infty}^{+\infty} \frac{1}{\sqrt{2\pi}} e^{-\frac{1}{2}z^2} dz = 1 . \tag{101}$$

Let $u = \frac{1}{\sqrt{2}} z$, then $du = \frac{1}{\sqrt{2}} dz$, and this expression becomes:

$$\frac{1}{\sqrt{\pi}} \int_{-\infty}^{+\infty} e^{-u^2} du = 1 \quad \rightarrow \quad \int_{-\infty}^{+\infty} e^{-u^2} du = \sqrt{\pi} \tag{102}$$

The integral on right side of the arrow is now of the form required by Gauss-Hermite quadrature with non-weight function g(u) from (6) equal to 1. Therefore, it can be estimated using Gauss-Hermite quadrature by:

$$\sqrt{\pi} = \int_{-\infty}^{+\infty} e^{-u^2} du \approx \sum_{j=1}^{12} w_j \, , \tag{103}$$

which is the LHS identity from (100). For the $2^{nd}$ identity in (100), consider two independent RVs $z_i \sim N(0,1)$, i=1, 2. Their joint PDF must also integrate to 1, thus:

$$\int_{-\infty}^{+\infty} \int_{-\infty}^{+\infty} \frac{1}{2\pi} e^{\left(-\frac{1}{2}z_1{}^2\right)} e^{\left(-\frac{1}{2}z_2{}^2\right)} dz_1 dz_2 = 1 \, . \tag{104}$$

Let $u_i = \frac{1}{\sqrt{2}} z_i$, then $du_i = \frac{1}{\sqrt{2}} dz_i$, for i=1, 2, so that (104) can be written as:

$$\pi = \int_{-\infty}^{+\infty} \int_{-\infty}^{+\infty} e^{(-u_1{}^2)} e^{(-u_2{}^2)} du_1 du_2 = \int_{-\infty}^{+\infty} e^{-u_1{}^2} du_1 \int_{-\infty}^{+\infty} e^{-u_2{}^2} du_2 \approx \sum_{i=1}^{12} \sum_{j=1}^{12} w_i w_j \, , \tag{105}$$

where the identity from (103) was applied twice.

As seen in Table 2, using an equity ratio of α=0.25, the best performing method is MGF(1) which uses the technique presented in this research with t ∈ {-1.0, -0.2}. When the equity ratio is α=0.50, the method presented in this research works best with t ∈ {-1.0, -0.2} for probabilities in the head and select upper tails, while t ∈ {-0.001, -0.005} works better for some probabilities in the center and upper tail of the distribution. Thus, if interest is in portfolios with equal weighting of stocks/bonds (i.e., α=0.50), an optimization technique such as that described by Mehta et al. (2007) would be beneficial using various combinations of t ∈ {$\tau_1, \tau_2$} along with some intuitive criteria or metric to determine the best performing t-set. Finally, with an equity ratio of α=0.75, the MGF(1) approach is generally more accurate in the head portion of the distribution, whereas MGF(2) is more accurate in the tail. These results again demonstrate the need to optimize over the 2-member t-set and determine which values perform best. We provide code to perform this optimization in Appendix B.

In terms of implementing these results using the code presented in Appendix A, we enter the lognormal parameters within the function main(), which is the application's entry point and exists within the code file LnSum.cpp. The following arrangement was used to derive the approximating lognormal mean and variance for MGF(2) with α=0.25 in Table 2.

```cpp
// Declare/initialize local variables.
//===================================
vector<double> uniMuVar;
long double tvals[2]={-0.001, -0.005};
```

```cpp
// Below are the variance-covariance matrix (V), the mean vector (m)
// and the sum constants (c) for the incoming lognormal random
// variables.  Change to V(3,3), M(3), and C(3) for a 3-term sum, etc...
//========================================================================
Eigen::MatrixXd V(2,2);
Eigen::VectorXd M(2),C(2);

// Set values for matrices and vectors.
//=====================================
V << 0.04635409, 0.00078, 0.00078, 0.00680625;
M << 1.0837, 1.0214;
C << 0.2500, 0.7500;
```

```cpp
// Invoke function to approximate a weighted lognormal sum.
//=========================================================
uniMuVar=LnSumApprox(M, V, C, tvals);
```

## VI.   Extension to a Sum of More than Two Terms

Consider the sum $S = a_1 \ddot{Y}_1 + a_2 \ddot{Y}_2 + \ldots + a_n \ddot{Y}_n$, where $a_i$ is a known constant and $\ddot{Y}_i \sim$ LogNormal$(\mu_{\ddot{y}_i}, \sigma_{\ddot{y}_i}^2)$ with Cov$(\ddot{Y}_i, \ddot{Y}_j) = \sigma_{(\ddot{y}_i, \ddot{y}_j)}$, for $i \neq j = 1, 2, \ldots, n$.  As with a 2-term sum, the distribution of S will be approximated with a univariate lognormal RV by solving the simultaneous equations in (77) and (78).  Regardless of how many terms constitute the sum, there will be two equations to solve for two unknown parameters.  The unknowns are the mean and variance of the normal distribution on which the approximating lognormal RV is based (using the scale factor).  In (77) and (78), only the constants $C_1$ and $C_2$ will change, and they represent two approximations to the MGF of S at different $t < 0$ values.  In vector notation,

$$S = \begin{pmatrix} a_1 & a_2 & \ldots & a_n \end{pmatrix} \begin{pmatrix} \ddot{Y}_1 \\ \ddot{Y}_2 \\ \vdots \\ \ddot{Y}_n \end{pmatrix}. \tag{106}$$

The expected value and variance of the sum S are known and given by:

$$E[S] = \begin{pmatrix} a_1 & a_2 & \ldots & a_n \end{pmatrix} \begin{pmatrix} \mu_{\ddot{y}_1} \\ \mu_{\ddot{y}_2} \\ \vdots \\ \mu_{\ddot{y}_n} \end{pmatrix}, \tag{107}$$

and,

$$V[S] = V\left[\begin{pmatrix} a_1 & a_2 & ... & a_n \end{pmatrix}\begin{pmatrix} \ddot{Y}_1 \\ \ddot{Y}_2 \\ \vdots \\ \ddot{Y}_n \end{pmatrix}\right] = \begin{pmatrix} a_1 & a_2 & ... & a_n \end{pmatrix} V\begin{pmatrix} \ddot{Y}_1 \\ \ddot{Y}_2 \\ \vdots \\ \ddot{Y}_n \end{pmatrix}\begin{pmatrix} a_1 \\ a_2 \\ \vdots \\ a_n \end{pmatrix} \quad (108)$$

$$= \begin{pmatrix} a_1 & a_2 & ... & a_n \end{pmatrix}\begin{bmatrix} \sigma^2_{\ddot{y}_1} & \sigma_{(\ddot{y}_1, \ddot{y}_2)} & ... & \sigma_{(\ddot{y}_1, \ddot{y}_n)} \\ \sigma_{(\ddot{y}_1, \ddot{y}_2)} & \sigma^2_{\ddot{y}_2} & ... & \sigma_{(\ddot{y}_2, \ddot{y}_n)} \\ \vdots & \vdots & \ddots & \vdots \\ \sigma_{(\ddot{y}_1, \ddot{y}_n)} & \sigma_{(\ddot{y}_2, \ddot{y}_n)} & ... & \sigma^2_{\ddot{y}_n} \end{bmatrix}\begin{pmatrix} a_1 \\ a_2 \\ \vdots \\ a_n \end{pmatrix}. \quad (109)$$

Here, $E[S]$ and $V[S]$ will be used to compute the starting points for Newton's method as they were in (88) and (89) for a 2-term sum. With n-terms, the constants $C_1$ and $C_2$ will consist of sums containing $12^n$ terms. To derive $C_1$ and $C_2$ we proceed exactly as in (36) for a 2-term sum. Here, let $\mathbf{\Sigma}$ be the variance-covariance matrix of the $u_i$'s, where,

$$u_i = \left(\frac{1}{\theta}\right)\ln(\ddot{y}_i) \text{ , for } i = 1, 2, ..., n \text{ ,} \quad (110)$$

are the underlying correlated normal RVs. Further, let $\mathbf{\Sigma} = \mathbf{LL}'$ be its Cholesky decomposition, where $\mathbf{L}$ is lower triangular, unique, and has positive real pivots. The transformation that decorrelates the PDF in (40) for a 2-term sum now becomes:

$$\text{Let } \vec{\boldsymbol{u}} = \sqrt{2}\mathbf{L}\vec{\boldsymbol{z}} + \vec{\boldsymbol{\mu}} \text{ , where } \vec{\boldsymbol{u}} = \begin{pmatrix} u_1 \\ u_2 \\ \vdots \\ u_n \end{pmatrix}, \ \vec{\boldsymbol{z}} = \begin{pmatrix} z_1 \\ z_2 \\ \vdots \\ z_n \end{pmatrix}, \ and, \ \vec{\boldsymbol{\mu}} = \begin{pmatrix} \mu_{x_1} \\ \mu_{x_2} \\ \vdots \\ \mu_{x_n} \end{pmatrix}. \quad (111)$$

Then,

$$\vec{\boldsymbol{z}} = \frac{1}{\sqrt{2}}\mathbf{L}^{-1}(\vec{\boldsymbol{u}} - \vec{\boldsymbol{\mu}}) . \quad (112)$$

To prove that this transformation is decorrelating in the $z_i$'s,

$$V(\vec{\boldsymbol{z}}) = \frac{1}{2}\mathbf{L}^{-1}V(\vec{\boldsymbol{u}} - \vec{\boldsymbol{\mu}})(\mathbf{L}^{-1})' = \frac{1}{2}\mathbf{L}^{-1}\mathbf{\Sigma}(\mathbf{L}')^{-1} = \frac{1}{2}\mathbf{L}^{-1}\mathbf{LL}'(\mathbf{L}')^{-1} = \frac{1}{2}\mathbf{I} . \quad (113)$$

Note that since L is lower-triangular with positive real pivots, it has the following general form:

$$\mathbf{L} = \begin{bmatrix} l_{11} & 0 & 0 & ... & 0 \\ l_{21} & l_{22} & 0 & ... & 0 \\ l_{31} & l_{32} & l_{33} & ... & 0 \\ \vdots & \vdots & \vdots & \ddots & \vdots \\ l_{n1} & l_{n2} & l_{n3} & ... & l_{nn} \end{bmatrix}, \tag{114}$$

where $l_{ii} > 0$, i=1, 2, …, n. This implies that $\overline{\boldsymbol{u}}$ from (111) consists of the following elements:

$$\begin{pmatrix} u_1 \\ u_2 \\ \vdots \\ u_n \end{pmatrix} = \begin{pmatrix} \sqrt{2}(l_{11}z_1) + \mu_{x_1} \\ \sqrt{2}(l_{21}z_1 + l_{22}z_2) + \mu_{x_2} \\ \vdots \\ \sqrt{2}(l_{n1}z_1 + l_{n2}z_2 + \cdots + l_{nn}z_n) + \mu_{x_n} \end{pmatrix}. \tag{115}$$

Here, $\ddot{Y}_i = 10^{X_i/10}$ where $X_i \sim N\left(\mu_{x_i}, \sigma_{x_i}^2\right)$, so that $\ddot{Y}_i$ follows the standard form lognormal distribution with underlying normal RVs $\theta X_i \sim N\left(\theta\mu_{x_i}, \theta^2\sigma_{x_i}^2\right)$, for i = 1, 2, …, n. By making this decorrelating transformation, we express the MGF of S in terms of the $z_i$'s as was done in (61) for a 2-term sum. The required weight functions appear for each $z_i$, i = 1, 2, …, n and the non-weight function in terms of the $\ddot{Y}_i$'s from (36) is now given by:

$$\left(\frac{1}{\pi}\right)^{\frac{n}{2}} e^{t(a_1\ddot{y}_1 + a_2\ddot{y}_2 + \ldots + a_n\ddot{y}_n)}. \tag{116}$$

In terms of the $u_i$'s, the non-weight function becomes:

$$\left(\frac{1}{\pi}\right)^{\frac{n}{2}} e^{t(a_1 e^{\theta u_1} + a_2 e^{\theta u_2} + \ldots + a_n e^{\theta u_n})}. \tag{117}$$

Finally, in terms of the $z_i$'s, the non-weight function is given by:

$$h(z_1, z_2, \ldots, z_n) = \tag{118}$$
$$\left(\frac{1}{\pi}\right)^{\frac{n}{2}} e^{t(a_1 e^{\theta(\sqrt{2}(l_{11}z_1)+\mu_{x_1})} + a_2 e^{\theta(\sqrt{2}(l_{21}z_1 + l_{22}z_2)+\mu_{x_2})} + \ldots + a_n e^{\theta(\sqrt{2}(l_{n1}z_1 + l_{n2}z_2 + \cdots + l_{nn}z_n)+\mu_{x_n})})}.$$

The constants $C_1$ and $C_2$ from (77) and (78) for an n-term sum S are then constructed by applying Gauss-Hermite quadrature to the n-dimensional integration of $h(\cdot)$ using two values for t < 0. As noted, the sum for each $C_i$ will consist of $12^n$ terms with each term representing a unique combination of the weight and root pairs across the n-dimensions. For example, the first term in the sum would use the first weight from Table 1 for each of the n-dimensions and each $z_i$ would be replaced by the corresponding first root from Table 1. This is repeated until all unique combinations have been represented. The weights are multiplied by each other as done on the left-hand sides of (75) and (76) for the 2-term case. The formal expression for $C_i$, i=1, 2 is:

$$C_i = \left(\frac{1}{\pi}\right)^{\frac{n}{2}} \sum_{k_1=1}^{12} \sum_{k_2=1}^{12} \cdots \sum_{k_n=1}^{12} \left[\left(\prod_{j=1}^{n} w_{k_j}\right) e^{t(a_1 e^{\theta(\sqrt{2}(l_{11}r_{k_1})+\mu_{x_1})} + a_2 e^{\theta(\sqrt{2}(l_{21}r_{k_1}+l_{22}r_{k_2})+\mu_{x_2})} +} \right. $$
$$\left. \cdots + a_n e^{\theta(\sqrt{2}(l_{n1}r_{k_1}+l_{n2}r_{k_2}+\cdots+l_{nn}r_{k_n})+\mu_{x_n}))}\right]. \tag{119}$$

Here, $(w_1, r_1)$ is the 1st weight/root pair from Table 1, $(w_2, r_2)$ is the 2nd, and so on.

## VII. Summary/Conclusion

Sums of lognormal RVs appear naturally in many disciplines and consequently must be modeled accurately within complex systems. Two common modeling procedures are the F-W (Fenton [1960]) and S-Y (Schwartz and Yeh [1982]) methods. Each has their benefits and drawbacks, for example, working well within some regions but not others. Mehta et al. (2007) propose a new and novel approach that is parameterizable, allowing the user to customize the CDF precision in regions of special interest. As is common with academic research, the paper assumes a high prerequisite level of technical expertise that may not be held by all who could benefit from it. We have therefore filled in the gaps and presented the material in a pedagogical fashion. We step the reader through all technical details required to understand the method for a (correlated) 2-term sum, and provide sufficient technical details for a full understanding of sums involving more than two (correlated) terms.

To emphasize the importance of such a procedure we provided an application to financial economics, and particularly to approximating CDF probabilities for the compounding return on a diversified portfolio of stocks and bonds, in discrete time. Such models are important within financial economics fields such as retirement planning where critical decisions on asset allocation and withdrawal rates are made periodically (e.g., yearly), not continuously. We have also included original source code from a C++ implementation that solves the required set of non-linear equations using Newton's method, with starting values motivated by the F-W approximation. Mehta et al. (2007) made use of MATLAB's built-in non-linear solvers. Such an implementation may suffer run-time inefficiencies within a large financial application being optimized over a planning period that spans several decades. Lower level programming languages can be more appropriate for such implementations. The technique we use converges rapidly, requiring only a small number of iterations.

As seen in Table 2, the improvements are modest when using arbitrary MGF values for t to generate the two required equations. Mehta et al. (2007) suggest that the user optimize over the t-set and find values suitable to their application. For example, CDF probabilities for a set of sum values would be generated via simulation over a region of interest, or the entire sum domain, as was done in Table 2. Optimization over the 2-member t-set would then compute the (weighted) sum of absolute %-deviations between the simulated and approximated values, and the best performing t-set would be chosen for that particular application. The best performing t-set would be the one that yields the minimum sum value. To achieve greater accuracy over particular regions of the lognormal sum domain, weights can be introduced for each absolute %-deviation (see, Mehta et al. [2007]).

Arguably, the F-W approach can be similarly optimized using various mean/variance combinations, but there is a distinction. Under the Mehta et al. (2007) framework, the optimal t-set may work well for a variety of related sums, whereas, the F-W optimization would need to be repeated whenever the sum changes. We have included C++ source code to simulate sum values and optimize the t-set in Appendix B. The computations are multi-threaded to reduce processing time. With respect to the finance application provided in Section V, we derived optimal t-sets and these are shown in Figure 2 along with the corresponding univariate lognormal CDF approximations. The sum of absolute %-differences was unweighted for this implementation which results in greater absolute precision in the head portion of the distribution, and this is clearly seen as $\alpha$ increases. We can enhance the univariate approximations shown in Figure 2 by weighting the sum of absolute %-differences in a manner that gives more importance to increasing values on the lognormal sum domain, or by partitioning the domain into sections and optimizing the t-set within each section. The univariate approximation would then be conditional on the section that a particular domain value resides in. Figure 3 shows the result of using a simple weighting scheme, namely, values of S < 0.75 receive a weight of 1.0, values between 0.75 and 1.10 receive a weight of 15.0, and values > 1.10 receive a weight of 50.0.

**Figure 2**
**CDF Approximations using the Unweighted Optimized t-Set**

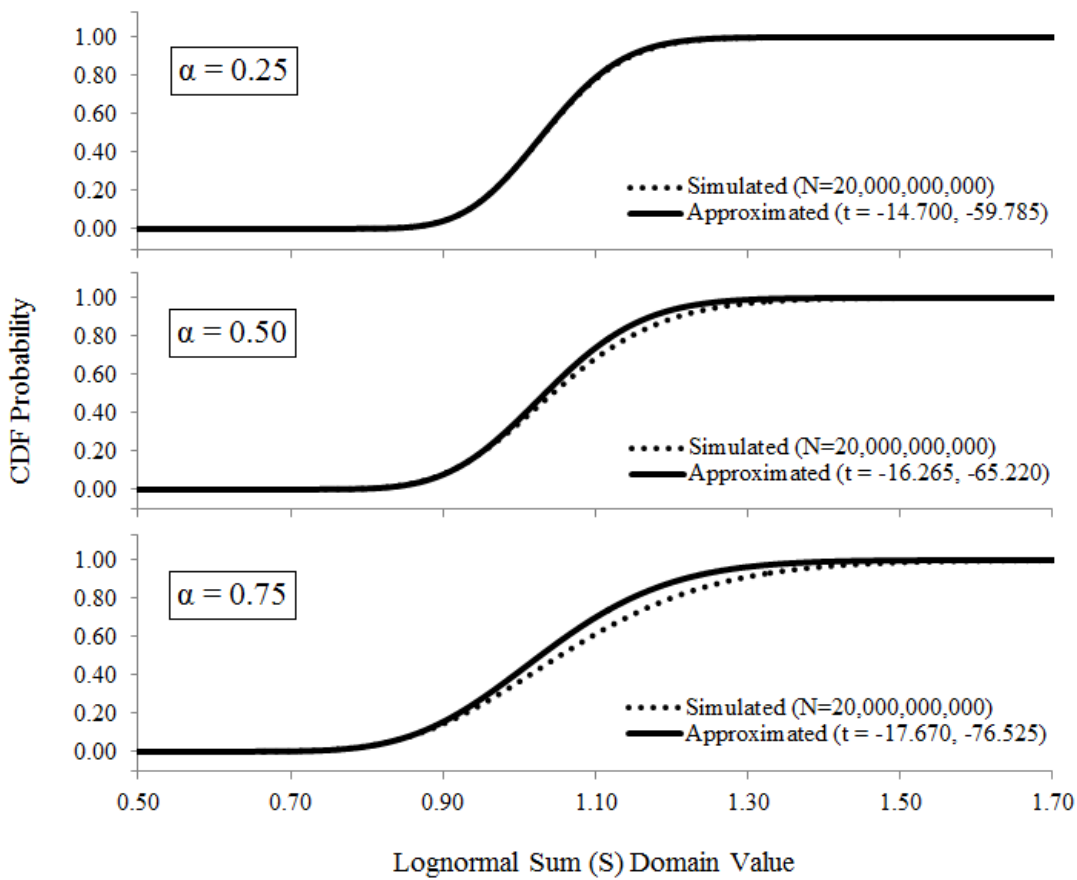

**Figure 3**
**CDF Approximations for α = 0.75 using the Weighted Optimized t-Set**

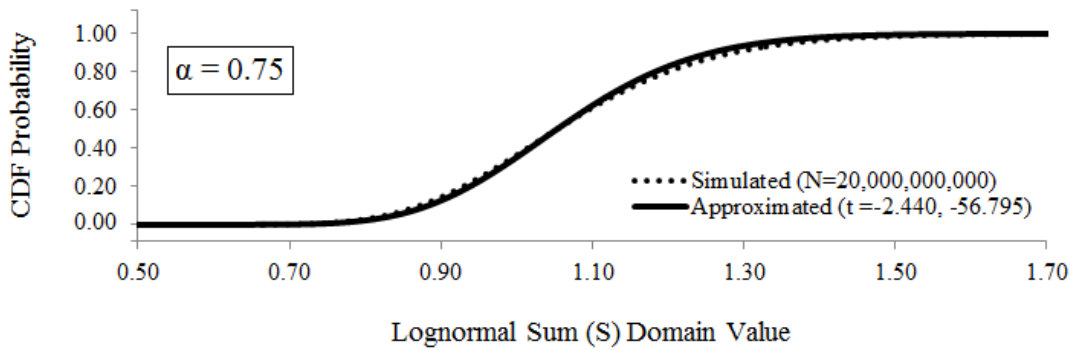

# Appendix A:  Lognormal Approximation Source Code

## Filename:   stdafx.h

```
/*
/ The MIT License (MIT)
/
/ Copyright (c) 2015 Chris Rook
/
/ Permission is hereby granted, free of charge, to any person obtaining a copy of this software and associated documentation files (the "Software"),
/ to deal in the Software without restriction, including without limitation the rights to use, copy, modify, merge, publish, distribute, sublicense,
/ and/or sell copies of the Software, and to permit persons to whom the Software is furnished to do so, subject to the following conditions:
/
/ The above copyright notice and this permission notice shall be included in all copies or substantial portions of the Software.
/
/ THE SOFTWARE IS PROVIDED "AS IS", WITHOUT WARRANTY OF ANY KIND, EXPRESS OR IMPLIED, INCLUDING BUT NOT LIMITED TO THE WARRANTIES OF MERCHANTABILITY,
/ FITNESS FOR A PARTICULAR PURPOSE AND NONINFRINGEMENT. IN NO EVENT SHALL THE AUTHORS OR COPYRIGHT HOLDERS BE LIABLE FOR ANY CLAIM, DAMAGES OR OTHER
/ LIABILITY, WHETHER IN AN ACTION OF CONTRACT, TORT OR OTHERWISE, ARISING FROM, OUT OF OR IN CONNECTION WITH THE SOFTWARE OR THE USE OR OTHER
/ DEALINGS IN THE SOFTWARE.   (License source: http://opensource.org/licenses/MIT)
/
/ Filename:  stdafx.h
/
/ Summary:
/
/    This is the header file where we include other header files, define constants, namespaces, inline functions, and function prototypes.
/
/**********************************************************************************************************************************************************/
#pragma once

// Include files.
//=================
#include "targetver.h"
#include <stdio.h>
#include <stdlib.h>
#include <iostream>
#include <Eigen/Dense>
#include <Eigen/Eigenvalues>
#include <Eigen/Cholesky>
#include <vector>
#include <algorithm>
#include <random>
#include <boost/thread/thread.hpp>
#include <boost/math/distributions.hpp>
#include <iostream>
#include <fstream>

using namespace std;
```

```cpp
// Constants.
//=============
const double pi = 3.141592653589793;
const double sf = log(10.00)/10.00;         // Scaling factor ln(10)/10 to align with Mehta et al. (2007)

// Inline functions to derive the underlying normal mean/variances from the lognormal mean/variances.
//=====================================================================================================
inline long double NMean(const Eigen::VectorXd inM, const Eigen::MatrixXd inV, const int i)
                   {return (1.0/sf)*(log(inM(i)) - (0.5)*log(1.0 + inV(i,i)/pow(inM(i),2)));}        // See (24).

inline long double NVar(const Eigen::VectorXd inM, const Eigen::MatrixXd inV, const int i, const int j)
                  {return pow((1.0/sf),2)*log(1.0 + inV(i,j)/abs(inM(i)*inM(j)));}                    // See (25) & (26).

// Function prototypes.
//=======================
vector<double> LnSumApprox(const Eigen::VectorXd inM, const Eigen::MatrixXd inV, const Eigen::VectorXd inC, const long double t[2]);
void GHQuad(const int curD, const long double tval, const Eigen::VectorXd inC, const long double allRts[12], const long double allWts[12], const
            Eigen::MatrixXd Ldc, const Eigen::VectorXd Mu, long double *uWts, long double *uRts, long double *rSum);
long double ProdTerm(const long double tval, const Eigen::VectorXd inC, const long double *nRts, const long double *nWts, const Eigen::MatrixXd inL,
                     const Eigen::VectorXd inM);
vector<long double> ThrdSimProb(const int inplproc, const long long int inn, const long int ink, const double * indvals, const Eigen::VectorXd inC,
                     const Eigen::VectorXd inNM, const Eigen::MatrixXd inV);
void SimProb(const int dim, const long long int simn, const long int ink, const double * indvals, long long int * CDFcnts, const Eigen::VectorXd inC,
             const Eigen::VectorXd inNM, const Eigen::MatrixXd inL);
vector<long double> ThrdtSetOpt(const int inplproc, const long int inul, const long int inprec, const long int ink, const Eigen::VectorXd inM, const
             Eigen::MatrixXd inV, const Eigen::VectorXd inC, const vector<long double> inCDFvals, const double * indvals);
void tSetOpt(const long int *inparms, const Eigen::VectorXd inM, const Eigen::MatrixXd inV, const Eigen::VectorXd inC, const vector<long double>
             inCDFvals, const double * indvals, long double *oVals);
```

# **Filename: LnSum.cpp**

```
/ Filename:  LnSum.cpp
/
/ Function:  main()
/
/ Summary:
```

```
/    The main() function here is the entry point for the application.  We are creating a function that, once compiled, can be invoked from within any
/    C++ application to approximate the distribution of a weighted sum of (correlated) lognormal random variables.  The method we are implementing is
/    from the research paper titled "Approximating a Sum of Random Variables with a Lognormal" by Mehta, Wu, Molisch, and Zhang (2007).  This paper
/    was published in the IEEE Transactions on Wireless Communications, Volume 6, Number 7.  Here, main() is used to create the quantities needed for
/    the function call, and then to invoke the function for testing.  Any user of this application will not need the main() function.  They will only
/    need the header file along with the 3 functions LnSumApprox(), GHQuad(), and ProdTerm().  Once compiled, LnSumApprox() can be invoked from
/    within their application.  The quantities we construct in main() are as follows:
/
/    1.) A vector of means for the lognormal random variables being summed (M).
/    2.) The variance-covariance matrix for the lognormal random variables being summed (V).
/    3.) A vector of the sum weights.  The first weight is for the first lognormal random variable, etc... (C).
/    4.) Settings for t from the moment generating function.  We have 2 unknowns therefore will need 2 equations to create 2 equations, regardless of
/        the number of lognormal random variables being summed.  (There is always just a single mean and variance for the approximating lognormal
/        distribution.)  This is the 2-member array tvals[].
/
/    These 4 quantities are passed as arguments to the function we create.  The function will approximate the distribution of the sum with a single
/    lognormal random variable.  It returns a 2-element vector with the mean and variance of the approximating lognormal random variable.  The user
/    then supplies these values as needed to a standard lognormal CDF/PDF call using built-in functions to derive probabilities for their sum.  The
/    function that derives the mean and variance for the approximating univariate lognormal random variable is named LnSumApprox().  Two other
/    functions are involved and are invoked from within LnSumApprox().  If the user invokes a normal CDF on the log domain then they can convert to
/    the corresponding normal mean and variance using the 2 inline functions NMean() and NVar() which are defined in the header file.
/***********************************************************************************************************************************************/
#include "stdafx.h"
int main(int argc, char *argv[])
{
        // Declare/initialize local variables.
        //==================================
        vector<double> uniMuVar;
        long double tvals[2]={-1.00, -0.20}, alpha=0.75;

        // Below are the variance-covariance matrix (V), the mean vector (m) and
        // the sum constants (c) for the incoming lognormal random variables.
        //======================================================================
        Eigen::MatrixXd V(2,2);
        Eigen::VectorXd M(2),C(2);

        // Set values for matrices and vectors.
        //=====================================
        V << 0.04635409, 0.00078, 0.00078, 0.00680625;
        M << 1.0837, 1.0214;
        C << alpha, 1-alpha;

        // Invoke function to approximate a weighted lognormal sum.
        //=========================================================
        uniMuVar=LnSumApprox(M, V, C, tvals);
        cout.setf(ios_base::fixed, ios_base::floatfield); cout.precision(20);
        cout << "Mean and variance of univariate approximating lognormal RV: " << endl;
        cout << "Mean =" << uniMuVar[0] << endl;
        cout << "Variance =" << uniMuVar[1] << endl << endl;
        cout << endl << "Done, hit return to exit." << endl; cin.get();
        return 0;
}
```

# Filename: LnSumApprox.cpp

```
/*
/ The MIT License (MIT)
/
/ Copyright (c) 2015 Chris Rook
/
/ Permission is hereby granted, free of charge, to any person obtaining a copy of this software and associated documentation files (the "Software"),
/ to deal in the Software without restriction, including without limitation the rights to use, copy, modify, merge, publish, distribute, sublicense,
/ and/or sell copies of the Software, and to permit persons to whom the Software is furnished to do so, subject to the following conditions:
/
/ The above copyright notice and this permission notice shall be included in all copies or substantial portions of the Software.
/
/ THE SOFTWARE IS PROVIDED "AS IS", WITHOUT WARRANTY OF ANY KIND, EXPRESS OR IMPLIED, INCLUDING BUT NOT LIMITED TO THE WARRANTIES OF MERCHANTABILITY,
/ FITNESS FOR A PARTICULAR PURPOSE AND NONINFRINGEMENT. IN NO EVENT SHALL THE AUTHORS OR COPYRIGHT HOLDERS BE LIABLE FOR ANY CLAIM, DAMAGES OR OTHER
/ LIABILITY, WHETHER IN AN ACTION OF CONTRACT, TORT OR OTHERWISE, ARISING FROM, OUT OF OR IN CONNECTION WITH THE SOFTWARE OR THE USE OR OTHER
/ DEALINGS IN THE SOFTWARE.  (License source: http://opensource.org/licenses/MIT)
/
/ Filename:  LnSumApprox.cpp
/
/ Function:  LnSumApprox()
/
/ Summary:
/
/    This function is invoked by a calling program and returns the univariate parameters for approximating a sum of (correlated) lognormal RVs.  The
/    input for this function is specified below and includes the means, variances, and covariances of the lognormal RVs that constitute the sum,
/    along with the constants that multiply these lognormal RVs while constructing the sum.  Also, the two values of t that form the moment-
/    generating function equations needed to obtain a solution for the 2 unknowns are provided.  These can be used to tune the approximating
/    lognormal RV to perform well in the tail or head portion based on the user's needs.  Regardless of how many lognormal RVs constitute the sum,
/    there will always be only 2 equations to solve for the underlying normal mean and variance which are converted to the mean and variance for the
/    approximating lognormal RV.  This function begins by declaring and initializing several local variables/objects including 2 arrays to hold the
/    weights and roots needed to implement Gauss-Hermite quadrature (assuming n=12).  Next, the means, variances, and covariances for the underlying
/    normal random variables are derived.  They are needed to structure the lognormal sum moment-generating function as needed by Gauss-Hermite
/    quadrature (i.e., with the proper weight function).  With the underlying variance-covariance matrix derived for the (correlated) normal random
/    variables we then confirm that it is positive definite by inspection of the smallest eigenvalue. (It must be > 0.)  Generally, we will assume
/    the underlying joint normal random variables are non-degenerate as this will allow the decorrelating transformation needed to put the moment-
/    generating function in an estimable form.  The next step is to perform the Cholesky decomposition of the variance-covariance matrix and this
/    yields the matrix necessary to perform the decorrelating transformation.  This decorrelating transformation is then applied to yield a specific
/    value for the moment-generating function of this sum by applying Gauss-Hermite quadrature to approximate the resulting integral.  Using 2 values
/    for t yields 2 constants, which we call C1 and C2.  Each is then set equal to the univariate moment-generating function with parameters for the
/    mean and variance of the underlying normal distribution to form the 2 equations solved for these 2 unknowns.  By subtracting the constants from
/    each equation we end up with a system of 2 non-linear functions equal to zero which are solved for the 2 unknown quantities using Newton's
/    method.  Here, we approximate the system with a 1st order Taylor series and solve the linear approximation for the unknowns.  These solutions
/    then become the starting points for the next iteration.  We iterate until both functions are within epsilon of zero.  Here, the convergence
/    criteria is taken as epsilon = (0.1)^10.  We start the procedure by taking the mean and variance for the underlying normal distribution that
/    would equal E[S] and V[S].  Since S is a sum of the form S = aY1 + bY2 + cY3 + etc ..., we know its mean and variance.  Thus, we start the
/    procedure by assuming that the best approximating univariate lognormal is the one with mean and variance equal to E[S] and V[S].  (This starting
/    point is motivated by the F-W procedure.)  Once the procedure ends, we have the preferred mean and variance from the underlying normal RVs and
/    these are then converted into the mean and variance of the preferred univariate approximator for the sum.  These 2 values are then returned to
/    the calling function in a 2-element vector.
/
/ Inputs:
```

```
/
/    1.) Vector of means for the lognormal random variables being summed.  If there are N lognormal random variables being summed then this vector
/         will contain N elements.
/    2.) Variance-covariance matrix for the lognormal random variables being summed.  If there are N lognormal random variables being summed then
/         this matrix will be NxN, symmetric, and positive definite.  The i-th diagonal term is the variance of random variable #i, and the ij-th term
/         is the covariance between random variable #i and #j.
/    3.) Vector of constants for the weighted sum.  The weighted sum is of the form:  S = aY1 + bY2 + cY3 + etc ..., and it has N terms.  Here the
/         constants are a, b, c, etc ...  If there are N terms involved in the sum then this vector has N terms.
/    4.) A 2-element array of values for t from the moment generating function definition.  Suggestions for good values to use in various scenarios
/         are provided by Mehta et al. (2007), who also suggest an optimization routine over the 2-member t-set.  If these values are identical then
/         the problem reduces to solving 1 non-linear equation with 2 unknowns.  Therefore, they should be distinct.
/
/ Outputs:
/
/    This function returns a 2-element vector containing the mean and variance for the approximating univariate lognormal random variable.  The
/    univariate lognormal random variable with this mean and variance approximates the distribution of the sum S.
/
/*************************************************************************************************************************************************/
#include "stdafx.h"
vector<double> LnSumApprox(const Eigen::VectorXd inM, const Eigen::MatrixXd inV, const Eigen::VectorXd inC, const long double t[2])
{
        // Get dimension of problem.
        //============================
        int n=(int) inM.size();

        // Declare/initialize local variables:
        //=====================================
        vector<double> uniPrms, uniLnPrms;                    // 2-element vectors for univariate normal and lognormal parameters for approx.
        Eigen::VectorXd nM(n);                                // Vector to hold underlying normal RVs means.
        Eigen::MatrixXd nV(n,n);                              // Matrix to hold underlying normal RVs variance-covariance matrix.
        Eigen::EigenSolver<Eigen::MatrixXd> esolver;          // Eigensolver object to hold eigenvalues.
        Eigen::MatrixXd L(n,n);                               // Matrix to hold left root of Cholesky decomposition.
        complex<double> egnval;                               // Variable to hold eigenvalues which, in general, can be complex.
        double minegnval;                                     // Variable to hold minimum eigenvalue.
        long double C[2], rSum[1],                            // 2-Dim array to hold LHS constant terms C1 and C2 and running sum total.
                     rts[12]={-0.314240376254, 0.314240376254,            /* Gauss-Hermite integration roots.  (See, Table 1.)   */
                              -0.947788391240, 0.947788391240,
                              -1.597682635153, 1.597682635153,
                              -2.279507080501, 2.279507080501,
                              -3.020637025121, 3.020637025121,
                              -3.889724897870, 3.889724897870},
                     wts[12]={0.570135236262500000, 0.570135236262500000,      /* Gauss-Hermite integration weights.  (See, Table 1.) */
                              0.260492310264200000, 0.260492310264200000,
                              0.051607985615880000, 0.051607985615880000,
                              0.003905390584629000, 0.003905390584629000,
                              0.000085736870435880, 0.000085736870435880,
                              0.000000265855168436, 0.000000265855168436};

        long double *uWts = new long double [n],              // Unique arrays of weights for single kernal value.
                    *uRts = new long double [n];              // Unique arrays of roots for single kernel value.

        // Issue a warning if the t-values are equal since one equation with 2 unknowns will be solved.
        //=============================================================================================
```

```cpp
if (t[0] == t[1])
{
    cout.setf(ios_base::fixed, ios_base::floatfield); cout.precision(20);
    cout << endl << "WARNING: The t-values are equal, t[0]=" << t[0] << " and t[1]=" << t[1] << "." << endl;
    cout         << "         As a result, one equation will be used to solve for 2 unknowns." << endl << endl;
}

// Parameters for the lognormal random variables have been provided.  Convert these
// to the corresponding parameters for the underlying normal random variables.
//=================================================================================
/* Normal RV means. */
for (int i=0; i<n; ++i)
    nM(i)=NMean(inM, inV, i);                   // See (24).

/* Normal RV variance-covariances. */
for (int i=0; i<n; ++i)
    for (int j=0; j<n; ++j)
        nV(i,j)=NVar(inM, inV, i, j);           // See (25) & (26).

// Get the eigenvalues of the normal RV variance-covariance matrix.
// (Eigenvectors are not needed, we will use the Cholesky decomposition to decorrelate.)
//=================================================================================
esolver.compute(nV, false);

// Get the minimum eigenvalue and confirm the covariance matrix is positive definite.
// Exit with an error if it is not.  We depend on this characteristic for the decomposition.
//=================================================================================
for (int i=0; i<n; ++i)
{
    egnval=esolver.eigenvalues()[i];
    if (i==0)
        minegnval=egnval.real();
    else
        minegnval=min(egnval.real(), minegnval);
}
if (minegnval <= 0.00)
{
    cout << "ERROR: Normal RV covariance matrix is not positive definite." << endl;
    cout << "EXITING...LnSumApprox()..." << endl; cin.get();
    exit (EXIT_FAILURE);
}

// Retrieve the Cholesky decomposition of the normal RV covariance matrix.
//=================================================================================
L = nV.llt().matrixL();                         // See (114).

// Build the LHS constants C1 and C2.  (Account for pi here.)
//=================================================================
for (int tt=0; tt<2; ++tt)
{
    GHQuad(n, t[tt], inC, rts, wts, L, nM, uWts, uRts, rSum);      // Constants C1 and C2, see (119).
    C[tt]=(1/pow(sqrt(pi),(int) n*1))*rSum[0];                     // See (119) for the general case and (77), (78) for the 2-term sum case.
}
```

```cpp
// Initial values for Newton's method will be those that correspond
// to the mean and variance of the sum being approximated.
//=====================================================================
Eigen::VectorXd ES(1);   ES(0)=inC.transpose()*inM;          // See (88).
Eigen::MatrixXd VS(1,1); VS(0)=inC.transpose()*inV*inC;       // See (89).

// Corresponding normal mean and standard deviation.
//===================================================
uniPrms.push_back(NMean(ES, VS));
uniPrms.push_back(sqrt(NVar(ES, VS, 0, 0)));

// Apply Newton's method to solve the system of 2 non-linear equations with 2 unknowns.
// Form system as Ax = b and solve for x, then back out the updated values for Mu & Sigma.
//=========================================================================================
Eigen::MatrixXd A(2,2);
Eigen::VectorXd b(2), updt(2);
long double s, mval;
int cont=1;

// Iterate using Newton's method to find a solution for the 2 non-linear
// equations and 2 unknowns.
//======================================================================
while (cont==1)
{
    // Populate vector b.
    //====================
    for (int i=0; i<2; ++i)
    {
        s = 0.00;
        for (int m=0; m<12; ++m)
            s = s + (wts[m]*exp(t[i]*exp(sf*(sqrt(2.0)*rts[m]*uniPrms[1] + uniPrms[0]))));    // See (77) & (78).
        b(i)=-((1/sqrt(pi))*(s) - C[i]);                                                      // See RHS of (82).
    }

    // Retrieve maximum abs value of function evaluated at current solution.
    //=====================================================================
    if (abs(b(0)) > abs(b(1)))
        mval=abs(b(0));
    else
        mval=abs(b(1));

    // Check for convergence (criteria is both equations < 0.0000000001).
    //=================================================================
    if (mval < pow(0.1,10.0))
        cont=0;

    // Populate matrix A using the starting values derived above for this iteration.  This only occurs when another iteration is required to
    // achieve convergence.  Then solve the equation and update the solution for another iteration.
    //====================================================================================================================================
    if (cont==1)
    {
        for (int i=0; i<2; ++i)  /* Column #1 */
```

```cpp
            {
                s = 0.00;
                for (int m=0; m<12; ++m)
                    s = s + (wts[m]*exp(t[i]*exp(sf*(sqrt(2.0)*rts[m]*uniPrms[1] + uniPrms[0])))*exp(sf*(sqrt(2.0)*rts[m]*uniPrms[1]
                        + uniPrms[0])));
                A(i,0)=(sf*t[i]/sqrt(pi))*s;                // See (84b).
            }
            for (int i=0; i<2; ++i)  /* Column #2 */
            {
                s = 0.00;
                for (int m=0; m<12; ++m)
                    s = s + (wts[m]*rts[m]*exp(t[i]*exp(sf*(sqrt(2.0)*rts[m]*uniPrms[1]
                          + uniPrms[0])))*exp(sf*(sqrt(2.0)*rts[m]*uniPrms[1] + uniPrms[0])));
                A(i,1)=(sf*t[i]*sqrt(2.00/pi))*s;           // See (85b).
            }

            // Update the solution.
            //========================
            updt=A.colPivHouseholderQr().solve(b);          // See (82).
            uniPrms[0]=updt(0)+uniPrms[0];                  // See (83).
            uniPrms[1]=updt(1)+uniPrms[1];                  // See (83).
        }
    }

    // Delete temporary memory allocations.
    //=====================================
    delete [] uWts;  uWts = nullptr;
    delete [] uRts;  uRts = nullptr;

    // Derive and return the corresponding mean and variance for the approximating univariate lognormal random variable.
    //====================================================================================================================
    uniLnPrms.push_back(exp(sf*uniPrms[0] + (0.5)*pow(sf*uniPrms[1],2.0)));                              // LN Approximating Mean.  See (21).
    uniLnPrms.push_back(exp(2.0*sf*uniPrms[0] + pow(sf*uniPrms[1],2.0))*(exp(pow(sf*uniPrms[1],2.0))-1.0));  // LN Approximating Var.  See (22).
    return uniLnPrms;
}
```

## Filename:  GHQuad.cpp

/ Filename:  GHQuad.cpp
/
/ Function:  GHQuad()
/
/ Summary:
/
/    The moment generating function of a sum of N-lognormal random variables is approximated using Gauss-Hermite quadrature.  Each dimension of the
/    sum is approximated with a 12-sum term reflecting the 12 weights and roots of the quadrature rule.  Therefore the N-integral moment-generating
/    function is replaced with a 12^N term sum.  If there are 2 dimensions in the sum (i.e., 2 lognormal RVs involved in the weighted sum) then the
/    double integral is approximated by a 12x12 = 144 term sum.  If there are 3 dimensions in the sum, then the moment-generating function's triple
/    integral is replaced by a 12^3 = 1728 term sum.  If there are 4 dimensions in the sum, then the moment-generating function's quadruple integral
/    is replaced by a 12^4 = 20,736 term sum.  Each of these sums then produces an actual estimate for either C1 or C2 depending on which of the two
/    t-values it is combined with.  For a given problem dimension, each summand is nothing more than a combination of the weights multiplied by the
/    function being integrated (without the weight function) having it's values replaced by the corresponding combination of the roots (that is, by
/    the roots that correspond to the specific weight combination).  Therefore, constructing the 2 moment-generating estimates is essentially a
/    combinatorial problem where we simply need to generate all possible combinations of the weight/root pairs for all N dimensions.  For each single
/    combination we compute the function, multiply it by the weights, and then add it to the running sum.  This suggests a recursive function, which
/    is how this function is constructed.  This function is passed the current dimension being constructed and it populates the N-dimensional weight
/    and root arrays then recursively invokes itself with the dimension reduced by 1.  (Recall that a dimension refers to a single lognormal RV being
/    summed.)  When at the first (inner most) dimension we have constructed a unique combination of the weights and roots across all N dimensions and
/    now we can construct the function and add it to the running sum.  The function ProdTerm() is invoked to create this single summand for every
/    unique combination of the weights/roots.  We therefore proceed recursively building a unique weight/root combination and once it is built we
/    apply it for a single summand term.  The recursive behavior allows us to cover all unique combinations for any dimension, however, note that
/    the number of terms grows exponentially thus there is a practical limit to the number of terms constituting the sum.
/
/ Inputs:
/
/    1.) An integer reflecting the current dimension being processed.
/    2.) One of the 2 t-values we will use to construct a single approximation of the moment-generating function.
/    3.) The vector of constants that make up the sum S = aY1 + bY2 + cY3 + ...  (Here, they are a, b, c, ...)
/    4.) Array of 12 Gauss-Hermite quadrature roots.
/    5.) Array of 12 Gauss-Hermite quadrature weights.
/    6.) The left-hand-side matrix L from the Cholesky decomposition used in the decorrelating transformation.
/    7.) The vector of means from the corresponding normal random variables.  The decorrelating transformation uses this vector.
/    8.) Pointer to an N-dimensional array of long doubles to hold the unique combination of weights.  Once populated we invoke ProdTerm() to
/        generate the single summand value.
/    9.) Pointer to an N-dimensional array of long doubles to hold the unique combination of roots.  Once populated we invoke ProdTerm() to generate
/        the single summand value.
/   10.) Pointer to a long double to hold the running sum of each term generated once a unique combination of weights and roots has been constructed.
/
/ Outputs:
/
/    This function produces no output but it does update the value of the running sum term (rSum) which is a pointer to a double and passed to this
/    function as a modifiable argument.
/
/***********************************************************************************************************************************************/
#include "stdafx.h"
void GHQuad(const int curD, const long double tval, const Eigen::VectorXd inC, const long double allRts[12], const long double allWts[12], const
            Eigen::MatrixXd Ldc, const Eigen::VectorXd Mu, long double *uWts, long double *uRts, long double *rSum)
{

```
        // Initialize running sum to zero at beginning of initial call.
        //===============================================================
        if (curD == Mu.size())
                rSum[0]=0.00;

        // Iterate over 12 GH weight/root pairs for each dimension of the problem.
        //=======================================================================
        for (int k=0; k<12; ++k)
        {
                // Set the unique weight/root pair for this dimension and index.
                //=============================================================
                uWts[curD-1]=allWts[k];     // See Table 1.
                uRts[curD-1]=allRts[k];     // See Table 1.

                // If not at first dimension continue building unique set of weights and roots.
                //============================================================================
                if (curD > 1)
                        GHQuad(curD-1, tval, inC, allRts, allWts, Ldc, Mu, uWts, uRts, rSum);
                else
                        rSum[0]=rSum[0]+ProdTerm(tval, inC, uRts, uWts, Ldc, Mu);   // See (119).
        }
}
```

# Filename:  ProdTerm.cpp

```
/ Filename:  ProdTerm.cpp
/
/ Function:  ProdTerm()
/
/ Summary:
/
/    An N-dimensional sum of correlated lognormals has its moment generating function approximated using 12-term Gauss-Hermite quadrature by a sum
/    consisting of 12^N terms.  Each single term represents a unique combination of the weight and root pairs.  The function GHQuad() invokes itself
/    recursively building these unique combinations and once a unique combination is built, this function is called to compute the single summand
```

```
/    term.  It is a product of many terms and is a function of the roots multiplied by the unique combination of weights.  The result is then
/    returned and added to the running sum.  Once all unique combinations of weights and roots have been accounted for the sum is complete and the
/    moment-generating function for a given t-value has been approximated using Gauss-Hermite quadrature.
/
/ Inputs:
/
/    1.) A 2-element array of values for t from the moment generating function definition.  Suggestions for good values to use in various scenarios
/        are provided by Mehta et al. (2007).
/    2.) Vector of constants for the weighted sum.  The weighted sum is of the form:  S = aY1 + bY2 + cY3 + etc ..., and it has N terms.  Here the
/        constants are a, b, c, etc ...  If there are N terms involved in the sum then this vector has N terms.
/    3.) Pointer to an N-dimensional array of long doubles to hold the unique combination of roots.
/    4.) Pointer to an N-dimensional array of long doubles to hold the unique combination of weights.
/    5.) The left-hand-side matrix L from the Cholesky decomposition used in the decorrelating transformation.
/    6.) The vector of means from the corresponding normal random variables.  The decorrelating transformation uses this vector.
/
/ Outputs:
/
/    This function generates a single output, which is a long double that represents the single product term for the summand which approximates the
/    moment-generating function of the sum S at a given t-value.  If the sum is N-dimensional, then there will be 12^N summands and this function is
/    invoked 12^N times generating 1 summand each time which are added together.
/
/*********************************************************************************************************************************************/
#include "stdafx.h"
long double ProdTerm(const long double tval, const Eigen::VectorXd inC, const long double *nRts, const long double *nWts, const Eigen::MatrixXd inL,
                     const Eigen::VectorXd inM)
{
        // Declare local variables.
        //===========================
        int dim=(int) inM.size();
        long double prduct=1.00, pt, et, dt, zt;

        // Create the single product term for a given set of t-values, constants, weights and roots.
        //=========================================================================================
        for (int i=0; i<dim; ++i)
        {
                // Build the kernel.
                //====================
                zt=0.0;
                for (int j=0; j<=i; ++j)
                        zt=zt+inL(i,j)*nRts[j];
                dt=sqrt(2.0)*zt+inM[i];
                et=exp(sf*dt);
                pt=exp(tval*inC[i]*et);
                prduct=prduct*nWts[i]*pt;      // Individual terms of (119).
        }

        // Return the product.
        //======================
        return prduct;
}
```

# Appendix B:  t-Set Optimization Source Code

## <u>Filename:</u>  LnSumOpt.cpp

```
/*
/ The MIT License (MIT)
/
/ Copyright (c) 2015 Chris Rook
/
/ Permission is hereby granted, free of charge, to any person obtaining a copy of this software and associated documentation files (the "Software"),
/ to deal in the Software without restriction, including without limitation the rights to use, copy, modify, merge, publish, distribute, sublicense,
/ and/or sell copies of the Software, and to permit persons to whom the Software is furnished to do so, subject to the following conditions:
/
/ The above copyright notice and this permission notice shall be included in all copies or substantial portions of the Software.
/
/ THE SOFTWARE IS PROVIDED "AS IS", WITHOUT WARRANTY OF ANY KIND, EXPRESS OR IMPLIED, INCLUDING BUT NOT LIMITED TO THE WARRANTIES OF MERCHANTABILITY,
/ FITNESS FOR A PARTICULAR PURPOSE AND NONINFRINGEMENT. IN NO EVENT SHALL THE AUTHORS OR COPYRIGHT HOLDERS BE LIABLE FOR ANY CLAIM, DAMAGES OR OTHER
/ LIABILITY, WHETHER IN AN ACTION OF CONTRACT, TORT OR OTHERWISE, ARISING FROM, OUT OF OR IN CONNECTION WITH THE SOFTWARE OR THE USE OR OTHER
/ DEALINGS IN THE SOFTWARE.  (License source: http://opensource.org/licenses/MIT)
/
/ Filename:  LnSumOpt.cpp
/
/ Function:  main()
/
/ Summary:
/
/    The main() function here is the entry point for the application that optimizes the univariate approximation over the 2-member t-set.  To find
/    the best performing t-set we first discretize the domain scale of the sum S = aY1 + bY2 + cY3 + etc ... into k values.  The CDF probability for
/    each value is then derived using simulation.  The k domain values and k CDF probabilities are stored in arrays/vectors.  We then iterate over
/    various 2-member combinations of t and, for each combination, we invoke the function LnSumApprox() to find the approximating univariate
/    lognormal mean and variance for this t-set.  Using these parameters we compute the approximated CDF probability and then take the (weighted)
/    absolute %-difference between the simulated and approximated values for each of the k domain values.  These k values are then summed and the
/    smallest sum of (weighted) absolute %-differences yields the optimal 2-member t-set.  As suggested by Mehta et al. (2007), weights can be added
/    to Taylor the approximation to suit the user's needs.  This can be done in the function tSetOpt() and we indicate where to add weights (if
/    desired) when constructing the sum.  The quantities we construct in main() are as follows:
/
/    1.) The # of independent processing units on the computer executing the optimization (plproc).
/    2.) The upper limit for both t1 and t2 which are integers that get converted to actual t-values (ul).  Raising this value increases the
/        granularity for t.  The variables t1 and t2 are iterated over to perform the optimization.
/    3.) The upper limit for the actual t-value (prec).
/    4.) The simulation sample size (n).
/    5.) The # of domain values for S on the log-scale to consider in the optimization (k).
/    6.) The array of k domain values for S (dvals).
/    7.) The vector of CDF values derived via simulation for these k domain values (CDFvals).
/    8.) The vector of optimal results (minimum sum of %-deviations, tvals[0], tvals[1], approximating univariate lognormal mean, approximating
/        univariate lognormal variance.
/
/    There are 4 functions involved in the t-set optimization, 2 for simulating the actual CDF probabilities of S = aY1 + bY2 + cY3 + etc ... at the k
/    domain values and 2 for performing the optimization.  Both sets of functions are split into (1) a wrapper that partitions the job and launches
/    individual parts concurrently in separate threads, and (2) the function that performs the task for a given set of inputs.  To simulate the CDF
/    probabilities for S at the k domain values set in dvals, ThrdSimProb() splits the task, multi-threads the call, and aggregates the results.  Each
```

```cpp
/   threaded call is executed by the function SimProb().  To optimize over the 2-member t-set the function ThrdtSetOpt() splits the job by sectioning
/   variable t1 into set sizes that increase in value.  Iteration over the 2-member t-set is then done section-by-section.  The results are inspected
/   and the best performing t-set across all calls to tSetOpt() is selected.  The results for the best performing t-set are returned by ThrdtSetOpt()
/   in a 5-member vector, then printed to the screen.
/
/***************************************************************************************************************************************************/
#include "stdafx.h"
int main(int argc, char *argv[])
{
        // Declare/initialize local variables.
        //===================================
        long double alpha=0.75;

        // Below are the variance-covariance matrix (V), the mean vector (m) and
        // the sum constants (c) for the incoming lognormal random variables.
        //=======================================================================
        Eigen::MatrixXd V(2,2);
        Eigen::VectorXd M(2),C(2);

        // Set values for matrices and vectors.
        //=====================================
        V << 0.04635409, 0.00078, 0.00078, 0.00680625;
        M << 1.0837, 1.0214;
        C << alpha, 1-alpha;

        // Find the # of independent processing units.
        //============================================
        int plproc = boost::thread::hardware_concurrency();
        cout << "Total # of threads: " << plproc << endl << endl;

        // Declare/initialize variables needed for t-set optimization.
        //===========================================================
        long int h=3, k=h*1000, ul=(long int)plproc*(1000/2), prec=100;      // Granularity when discretizing the sum domain. Upper limit for t1, t2
                                                                             // and corresponding divisor.
        long long int n=20000000000;                                         // Total simulation sample size.
        double *dvals = new double[k];                                       // High value on the sum domain, and array to hold sum domain values.
        long double *probs = new long double[k];                             // Array to hold the simulated probabilities for comparison.
        vector<long double> CDFvals, gOptVals;                               // Vector to hold simulated CDF probabilities for each domain value and
                                                                             // globally optimal t-set.

        // Populate array dvals with the domain values used for optimizing and simulate the corresponding probability.
        //===========================================================================================================
        for (int i=0; i<k; ++i)
                dvals[i]=(double) ((double) (i+1))*((double) h/(double) k);

        // Simulate the CDF probability for each domain value.
        //===================================================
        CDFvals=ThrdSimProb(plproc, n, k, dvals, C, M, V);

        // Find the optimal t-set.
        //========================
        gOptVals=ThrdtSetOpt(plproc, ul, prec, k, M, V, C, CDFvals, dvals);
```

```cpp
        // Output the final values
        //========================================
        cout.setf(ios_base::fixed, ios_base::floatfield); cout.precision(20);
        cout << endl << "Optimization Complete: " << endl;
        cout << "Optimal sum %-diff is gOptVals[0]=" << gOptVals[0] << endl;
        cout << "Optimal tval[0] is gOptVals[1]=" << gOptVals[1] << endl;
        cout << "Optimal tval[1] is gOptVals[2]=" << gOptVals[2] << endl;
        cout << "Univariate LN Mean is gOptVals[3]=" << gOptVals[3] << endl;
        cout << "Univariate LN Variance is gOptVals[4]=" << gOptVals[4] << endl;

        // Delete temporary memory allocations.
        //========================================
        delete [] dvals;  dvals = nullptr;
        delete [] probs;  probs = nullptr;

        cout << endl << "Done, hit return to exit." << endl; cin.get();
        return 0;
}
```

## Filename:   ThrdSimProb.cpp

```cpp
/*
/ The MIT License (MIT)
/
/ Copyright (c) 2015 Chris Rook
/
/ Permission is hereby granted, free of charge, to any person obtaining a copy of this software and associated documentation files (the "Software"),
/ to deal in the Software without restriction, including without limitation the rights to use, copy, modify, merge, publish, distribute, sublicense,
/ and/or sell copies of the Software, and to permit persons to whom the Software is furnished to do so, subject to the following conditions:
/
/ The above copyright notice and this permission notice shall be included in all copies or substantial portions of the Software.
/
/ THE SOFTWARE IS PROVIDED "AS IS", WITHOUT WARRANTY OF ANY KIND, EXPRESS OR IMPLIED, INCLUDING BUT NOT LIMITED TO THE WARRANTIES OF MERCHANTABILITY,
/ FITNESS FOR A PARTICULAR PURPOSE AND NONINFRINGEMENT. IN NO EVENT SHALL THE AUTHORS OR COPYRIGHT HOLDERS BE LIABLE FOR ANY CLAIM, DAMAGES OR OTHER
/ LIABILITY, WHETHER IN AN ACTION OF CONTRACT, TORT OR OTHERWISE, ARISING FROM, OUT OF OR IN CONNECTION WITH THE SOFTWARE OR THE USE OR OTHER
/ DEALINGS IN THE SOFTWARE.  (License source: http://opensource.org/licenses/MIT)
/
/ Filename:  ThrdSimProb.cpp
/
/ Function:  ThrdSimProb()
/
/ Summary:
/
/    This function splits the job of simulating CDF probabilities for the sum S = aY1 + bY2 + cY3 + etc ... across multiple threads to exploit the
/    computer's full capacity and reduce runtimes.  To simulate values from the joint distribution of (Y1, Y2, Y3, ...) we simulate values on the
/    corresponding joint distribution of (X1, X2, X3, ...), where Yi = exp(sf*Xi).  The Xi's are the corresponding underlying normal RVs and they
/    will be correlated when the Yi's are correlated.  To simulate observations on the Xi's we first decorrelate them to independent standard normal
/    RVs and simulate values on these RVs.  The independent standard normal RVs (Z1, Z2, Z3, ...) are then recorrelated to X values, which are then
/    transformed to Y values.  The simulated observation on the joint distribution of the Yi's is then used to construct a single simulated value for
/    S.  In this function we accept all parameters for the Yi's and derive the parameters for the underlying (correlated) normal random vector (X1,
/    X2, X3, ...).  We then derive the left root of the Cholesky decomposition used to decorrelate the normal random variables.  The parameters for
/    the normal random variables are then passed to the function SimProb() which will use this information to simulate values on S.  In this function
```

```
/     we split the simulation task into equal sized components and launch parallel jobs in separate threads.  Each thread will populate a single array
/     of CDF counts that we create in this function and pass as an argument to SimProb().  Once each of these arrays has been populated we combine the
/     counts and then derive the CDF probabilities as the total number of observations <= each domain value divided by the total N.  The vector of CDF
/     probabilities for each domain value is then returned by this function.
/
/ Inputs:
/
/     1.) The number of independent processing units on the computer running the program.
/     2.) The total simulation sample size for deriving CDF probabilities for the sum S = aY1 + bY2 + cY3 + etc ...
/     3.) The number of domain values for S that we will derive CDF probabilities for.
/     4.) The array of domain values for S.
/     5.) The vector of sum constants (a, b, c, ...).
/     6.) The vector of means for the lognormal RVs Y1, Y2, Y3, ...
/     7.) The variance-covariance matrix for the lognormal RVs Y1, Y2, Y3, ...
/
/ Output:
/
/     This function returns a vector with the simulated CDF probabilities for each domain value chosen for S.
/
/*****************************************************************************************************************************************/
#include "stdafx.h"
vector<long double> ThrdSimProb(const int inplproc, const long long int inn, const long long int ink, const double * indvals, const Eigen::VectorXd inC,
                                const Eigen::VectorXd inM, const Eigen::MatrixXd inV)
{
        // Get dimension of the problem.
        //===============================
        int dim = (int) inM.size();

        // Derive the corresponding vector of normal means.
        //=================================================
        Eigen::VectorXd nM(dim);
        for (int i=0; i<dim; ++i)
                nM(i)=NMean(inM, inV, i);

        // Derive the corresponding normal variance-covariance matrix.
        //============================================================
        Eigen::MatrixXd nV(dim,dim);
        for (int i=0; i<dim; ++i)
                for (int j=0; j<dim; ++j)
                        nV(i,j)=NVar(inM, inV, i, j);

        // Derive the Cholesky factorization of the variance-covariance matrix.
        //====================================================================
        Eigen::MatrixXd L(dim,dim);
        L = nV.llt().matrixL();

        // Split the total simulation sample size by # threads.
        //=====================================================
        long long int rsize=(long long int) inn/inplproc;

        // Need an array of CDF count arrays.  One per pllproc value.  These are updated in SimProb().
        //===========================================================================================
        long long int **CDFarys;
```

```cpp
CDFarys = new long long int * [inplproc];

// Construct a dynamically sized array of thread objects.
//=========================================================
boost::thread * t=new boost::thread[inplproc];

// Thread the simulations to exploit the machine's full capacity.
//================================================================
for (int p=0; p<inplproc; ++p)
{
        // Initialize the CDF counter array for this thread.
        //==================================================
        CDFarys[p] = new long long int [ink];

        // Launch a call to SimProb() for each thread.
        //============================================
        t[p] = boost::thread(SimProb, dim, rsize, ink, boost::cref(indvals), boost::ref(CDFarys[p]), inC, nM, L);
}

// Pause until all threads finish.
//================================
for (int i=0; i<inplproc; ++i)
        t[i].join();

// Accumulate the CDF counts across threads into the vector cVec, which is initialized to all zeros.
//=================================================================================================
vector<long long int> cVec;
for (int i=0; i<ink; ++i)
        cVec.push_back(0);
for (int i=0; i<inplproc; ++i)
        for (int k=0; k<ink; ++k)
                cVec[k] = cVec[k] + CDFarys[i][k];

// Delete temporary memory allocations.
//=====================================
for (int p=0; p<inplproc; ++p)
{
        delete [] CDFarys[p];  CDFarys[p]=nullptr;
}
delete [] CDFarys;  CDFarys=nullptr;
delete [] t;  t=nullptr;

// Compute the CDF probabilities.
//===============================
vector<long double> rVec;
for (int k=0; k<ink; ++k)
        rVec.push_back (((long double) cVec[k])/(rsize*inplproc));

// Return a vector with the CDF probabilities.
//============================================
return rVec;
}
```

# Filename: **SimProb.cpp**

```
/ Filename:  SimProb.cpp
/
/ Function:  SimProb()
/
/ Summary:
/
/    This function simulates the CDF probabilities for the correlated lognormal sum S = aY1 + bY2 + cY3 + etc ...  It does so by first finding the
/    parameters for the underlying correlated normal random variables (via transformation), and then by decorrelating those normal random variables
/    (via transformation).  Once the elements of these 2 transformations have been defined we simulate independent standard normal random variables
/    and apply the same 2 transformations in reverse order to produce observations on the joint distribution of (Y1, Y2, Y3, ...).  Once random
/    observations on the joint distribution have been generated we construct the simulated value for S.  For each element in the discretized domain
/    of S we compute the CDF probability as the number of simulated observations that are <= to it, divided by the total simulation sample size.
/    This function populates an array that is provided to it with only the count of observations that are <= each domain value.  These are referred
/    to as the CDF counts and the calling function uses this array to compute the CDF probabilities.
/
/ Inputs:
/
/    1.) Dimension of the problem, i.e. the number of terms involved in the sum S.
/    2.) The simulation sample size.
/    3.) The number of discretized domain values of S to compute the CDF counts for.
/    4.) The array of discretized domain values of S.  We compute a simulated CDF probability for each value.  (This array is of size specified in
/        parameter #3.)
/    5.) An empty array with the same size as the array in parameter #4 to hold the simulated CDF counts for each domain value of S.  This array will
/        be populated by the function.
/    6.) The vector of sum constants (a, b, c, ...).
/    7.) The vector of means for the lognormal RVs Y1, Y2, Y3, ...
/    8.) The Cholesky decomposition (left root) for the variance-covariance matrix of the underlying normal RVs.  In general the user will provide
/        the parameters for the lognormal RVs, including the variance-covariance matrix.  A transformation converts these to normal random variables
/        and their variance-covariance matrix is constructed using the lognormal variance-covariance matrix.  This parameter is the left root of the
/        decomposition for that matrix.
/
/ Output:
/
/    This function does not return a value.  It populates an empty array that is provided to it via parameter #5.  These are the CDF counts for each
```

```
/    discretized domain value of the sum S.
/
/*********************************************************************************************************************************/
#include "stdafx.h"
void SimProb(const int dim, const long long int simn, const long long int ink, const double * indvals, long long int * CDFcnts, const Eigen::VectorXd inC,
             const Eigen::VectorXd inNM, const Eigen::MatrixXd inL)
{
    // Declare/initialize needed local variables.
    //==========================================
    std::random_device rd;
    std::default_random_engine gen(rd());
    Eigen::VectorXd z(dim), u(dim), y(dim);
    double S;

    // Initialize CDF counter array to zero for each domain value of S.
    //================================================================
    for (int i=0; i<ink; ++i)
            CDFcnts[i]=0;

    // Simulate specified # of values for the sum S, which requires several steps as detailed below.
    //=============================================================================================
    for (int i=0; i<simn; ++i)
    {
            // Step #1:  Generate independent standard normal RVs.
            //===================================================
            for (int s=0; s<dim; ++s)
                    z(s)=std::normal_distribution<double> (0.0,1.0)(gen);

            // Step #2:  Correlate the independent standard normal RVs.
            //========================================================
            u=inL*z + inNM;

            // Step #3:  Transform back to correlated lognormal RVs.
            //=====================================================
            for (int l=0; l<dim; ++l)
                    y(l)=exp(sf*u(l));

            // Step #4:  Construct simulated value for the sum S.
            //==================================================
            S = inC.transpose()*y;

            // Update the CDF counter for each domain value.
            //=============================================
            for (int d=0; d<ink; ++d)
                    if (S <= indvals[d])
                            CDFcnts[d]=CDFcnts[d]+1;
    }
}
```

# Filename:  ThrdtSetOpt.cpp

/
/ Filename:  ThrdtSetOpt.cpp
/
/ Function:  ThrdtSetOpt()
/
/ Summary:
/
/    This function breaks up the job of optimizing the 2-member t-set and submits calls to tSetOpt() concurrently in separate threads to speed up
/    processing time.  The function tSetOpt() iterates over t-values and finds the one which minimizes the sum of (weighted) absolute %-deviations
/    from the simulated values.  In this application we do not apply a weight function and treat each absolute %-difference equally, but we do
/    indicate where to apply a weight function in the function tSetOpt().  A better solution can be obtained by applying a weight function, and it
/    will be specific to the user's application.  This function breaks up the t1 range into collections that increase in size by the variable "binc"
/    (bucket increment).  For example, if the executing computer has 20 independent processing units then the t1 values will be split into 20 sets
/    that increase in size by 20/2 = 10.  That is, set #2 has size equal to the size of set #1 + 10.  This is done because t2 ranges from t1+1 to the
/    upper range as set via inparms[2].  Therefore lower values of t1 have longer run times since they will process more values of t2.  This is the
/    justification for increasing the set size as t1 increases.  (The 2 MGF equations we generate using different values for t are the same for the
/    t-sets (x,y) and (y,x), and only one set needs to be evaluated.)  After all jobs are launched concurrently the program pauses until all threads
/    finish then inspects each oVals[] array for the globally optimal solution.  The oVals[] array holds the locally optimal settings for an
/    individual thread.  This function then returns a vector with the 5-member array of globally optimal settings.
/
/ Inputs:
/
/    1.) The number of independent processing units on the computer running the program.
/    2.) The upper limit on t1 and t2, which are integer values that get converted to actual MGF t-values.  This setting determines the granularity
/        of the actual t-values evaluated during the optimization.
/    3.) The maximum t-value.  For example, if this parameter is set to 100 then the largest actual t-value assessed during the optimization is 100.
/    4.) The number of domain values for S on the log-scale that we will consider.
/    5.) The vector of means for the lognormal RVs Y1, Y2, Y3, ...
/    6.) The variance-covariance matrix for the lognormal RVs Y1, Y2, Y3, ...
/    7.) The vector of sum constants (a, b, c, ...).
/    8.) The array of CDF values of size equal to the setting of parameter #4.
/    9.) The array of domain values of size equal to the setting of parameter #4.
/
/ Output:

```
/
/    A vector containing the following 5 quantities:
/       1.) The minimum sum of percent differences.
/       2.) The globally optimal setting for tvals[0].
/       3.) The globally optimal setting for tvals[1].
/       4.) The globally optimal approximating univariate lognormal mean.
/       5.) The globally optimal approximating univariate lognormal variance.
/
/**********************************************************************************************************************************/
#include "stdafx.h"
vector<long double> ThrdtSetOpt(const int inplproc, const long int inul, const long int inprec, const long int ink, const Eigen::VectorXd inM,
                                const Eigen::MatrixXd inV, const Eigen::VectorXd inC, const vector<long double> inCDFvals, const double * indvals)
{
        // Optimize over the 2-member t-set.
        //===================================
        int binc = (int) inplproc/2;                                    // Bucket increment size for t1.
        long int s = (long int) ((inul - (binc/2.00)*(inplproc-1)*(inplproc))/(inplproc)), start, stop=0;   // Variable s is the 1st bkt size for t1.
        long double **oVals = new long double * [inplproc];             // Array of arrays to hold the 5 optimal values for each run.
        long int **parms = new long int * [inplproc];                   // Array of arrays to hold the parameter settings which change per run.
        boost::thread * t=new boost::thread[inplproc];                  // Array of thread objects.

        // Optimization settings may not work.  Check here and exit with appropriate message.
        //===============================================================================
        if (s < 1)
        {
                cout << "ERROR: These optimization settings will not work.  Either:" << endl;
                cout << "       1.) Decrease the bucket increment size (binc)." << endl;
                cout << "       2.) Increase the upper limit (ul) to generate more t1 values." << endl;
                cout << "EXITING...ThrdtSetOpt()..." << endl; cin.get();
                exit (EXIT_FAILURE);
        }

        for (int p=0; p<inplproc; ++p)
        {
                // Set values required for this candidate set.
                //==========================================
                parms[p] = new long int [6];
                parms[p][2]=inul; parms[p][3]=inprec; parms[p][4]=ink; parms[p][5]=p+1;   // These parameter settings do not change.
                start = stop + 1;
                stop = start + (s + binc*p - 1);
                parms[p][0]=start; parms[p][1]=stop;                                     // These values do change.

                // Array to hold the locally optimal t-set and sum of %-diffs.
                //==========================================================
                oVals[p] = new long double [5];

                // Launch individual calls for sets of t1-values in separate threads to speed processing.
                //================================================================================
                t[p] = boost::thread(tSetOpt, boost::cref(parms[p]), inM, inV, inC, boost::cref(inCDFvals), boost::cref(indvals), boost::ref(oVals[p]));
        }
```

```cpp
        // Pause until all threads finish.
        //================================
        for (int p=0; p<inplproc; ++p)
                t[p].join();

        // Find the globally optimal t-set from the local optimums.
        //========================================================
        vector<long double> gVals;
        for (int p=0; p<inplproc; ++p)
        {
                if (p==0)
                {
                        for (int i=0; i<5; ++i)
                                gVals.push_back(oVals[p][i]);
                }
                else if (oVals[p][0] < gVals[0])
                {
                        gVals[0] = oVals[p][0];                         // Optimal sum of %-diffs value
                        gVals[1] = oVals[p][1]; gVals[2] = oVals[p][2];     // Optimal t-vals
                        gVals[3] = oVals[p][3]; gVals[4] = oVals[p][4];     // Corresponding univariate LN mean/variance
                }
        }

        // Free up dynamic memory allocations.
        //====================================
        for (int p=0; p<inplproc; ++p)
        {
                delete [] oVals[p]; oVals[p] = nullptr;
                delete [] parms[p]; parms[p] = nullptr;
        }
        delete [] oVals;  oVals = nullptr;
        delete [] parms;  parms = nullptr;
        delete [] t;   t = nullptr;

        // Return the optimal t-set along with the minimum sum %-diff and approximating univariate LN mean/variance.
        //===========================================================================================================
        return gVals;
}
```

```
/*
/ The MIT License (MIT)
/
/ Copyright (c) 2015 Chris Rook
/
/ Permission is hereby granted, free of charge, to any person obtaining a copy of this software and associated documentation files (the "Software"),
/ to deal in the Software without restriction, including without limitation the rights to use, copy, modify, merge, publish, distribute, sublicense,
/ and/or sell copies of the Software, and to permit persons to whom the Software is furnished to do so, subject to the following conditions:
/
/ The above copyright notice and this permission notice shall be included in all copies or substantial portions of the Software.
/
/ THE SOFTWARE IS PROVIDED "AS IS", WITHOUT WARRANTY OF ANY KIND, EXPRESS OR IMPLIED, INCLUDING BUT NOT LIMITED TO THE WARRANTIES OF MERCHANTABILITY,
/ FITNESS FOR A PARTICULAR PURPOSE AND NONINFRINGEMENT. IN NO EVENT SHALL THE AUTHORS OR COPYRIGHT HOLDERS BE LIABLE FOR ANY CLAIM, DAMAGES OR OTHER
/ LIABILITY, WHETHER IN AN ACTION OF CONTRACT, TORT OR OTHERWISE, ARISING FROM, OUT OF OR IN CONNECTION WITH THE SOFTWARE OR THE USE OR OTHER
/ DEALINGS IN THE SOFTWARE.  (License source: http://opensource.org/licenses/MIT)
/
/ Filename:  tSetOpt.cpp
/
/ Function:  tSetOpt()
/
/ Summary:
/
/    This function finds the optimal t-set pair for a given range of t1 and t2 values.  The variables t1 and t2 are integers and we iterate over each
/    computing a tval[0] and tval[1] based on the precision specified.  The optimal pair is the t-set that yields the lowest sum of (weighted)
/    absolute %-deviations between the approximating univariate lognormal CDF probabilities and the simulated CDF probabilities for S.  In this
/    application the sum of absolute %-deviations is unweighted which yields more absolute accuracy in the head portion of the distribution.  This
/    can be balanced by adding weights to the sum of absolute %-deviations (see for example, Mehta et al. (2007)).  This function populates a 5-
/    element array passed to it with the smallest sum of percentage deviations, along with the values tval[0] and tval[1] that yield the smallest
/    sum, and the approximating univariate lognormal's mean and variance.  This function is created to allow the optimization over the 2-member t-set
/    to be split amongst threads and launched in parallel to reduce runtime.
/
/ Inputs:
/
/    1.) A 6-member array of constants needed for this run.  The members of this array are:
/            inparms[0] = t1 start value for this iteration.  t1 is an integer and tval[0] is derived from it.
/            inparms[1] = t1 end value for this iteration.
/            inparms[2] = Upper limit for both t1 and t2.
/            inparms[3] = The maximum for tvals[0] and tvals[1].  We derive tval[i] as -ti/(inparms[2]/inparms[3]), i=1,2.  To increase the granularity
/                         of tvals[i], increase the value of inparms[2].
/            inparms[4] = The number of domain values for S on the log-scale that we will consider.
/            inparms[5] = The thread number.
/    2.) The vector of means for the lognormal RVs Y1, Y2, Y3, ...
/    3.) The variance-covariance matrix for the lognormal RVs Y1, Y2, Y3, ...
/    4.) The vector of sum constants (a, b, c, ...).
/    5.) The array of CDF values of size equal to inparms[4].
/    6.) The array of domain values of size equal to inparms[4].
/    7.) A 5-element empty array to hold the following items:
/            oVals[0] = Minimum sum of unweighted percent differences for this range of t1 values.
/            oVals[1] = Optimal tval[0] for this range of t1 values.
```

```
/            oVals[2] = Optimal tval[1] for this range of t1 values.
/            oVals[3] = The corresponding optimal approximating univariate lognormal mean for this range of t1 values.
/            oVals[4] = The corresponding optimal approximating univariate lognormal variance for this range of t1 values.
/
/ Output:
/
/    This function does not return a value.  It populates the empty array specified in parameter #7 with the optimal settings for this range of t1
/    values.
/
/******************************************************************************************************************************************************/
#include "stdafx.h"
void tSetOpt(const long int *inparms, const Eigen::VectorXd inM, const Eigen::MatrixXd inV, const Eigen::VectorXd inC, const vector<long double>
             inCDFvals, const double *indvals, long double *oVals)
{
        // Declare/initialize local variables.
        //======================================
        long double tvals[2];                  // Optimal %-diff sum and 2-member t-set.
        Eigen::VectorXd lnm(1);                 // Vector with one element for the LN mean.
        Eigen::MatrixXd lnv(1,1);               // Matrix with one element for the LN variance.
        boost::math::normal normdist;           // Distribution object for CDF call.
        vector<double> uniMuVar;                // Vector to hold LN mean/variance.
        oVals[0]=pow(10.0,10);                  // Sum accumulator, initialize to high initial value.

        // Iterate over the set of t-values provided and find the optimal one.
        //===================================================================
        for (int t1=inparms[0]; t1<=inparms[1]; ++t1)
              for (int t2=t1+1; t2<=inparms[2]; ++t2)
              {
                     // Define the 2-member t-set.
                     //============================
                     tvals[0]=(long double) -t1/(inparms[2]/inparms[3]);
                     tvals[1]=(long double) -t2/(inparms[2]/inparms[3]);

                     // Find the approximating lognormal for S.
                     //========================================
                     uniMuVar=LnSumApprox(inM, inV, inC, tvals);

                     // Use the normal distribution CDF for lognormal CDF values.
                     //==========================================================
                     lnm(0)=uniMuVar[0];  lnv(0)=uniMuVar[1];
                     normdist = boost::math::normal(sf*NMean(lnm, lnv, 0), sqrt(pow(sf,2)*NVar(lnm, lnv, 0, 0)));

                     // Compute the sum of all %-diffs with simulated CDF values.
                     //==========================================================
                     long double spd=0.0;
                     for (int i=0; i<inparms[4]; ++i)
                           if (inCDFvals[i] > 0.0)
                                  spd = spd + /* Add weight function here. */ abs(cdf(normdist, log(indvals[i])) - inCDFvals[i])/inCDFvals[i];

                     // Update optimal t-set.
                     //=======================
                     if (spd < oVals[0])
                     {
```
55

```
                oVals[0]=spd;
                oVals[1]=tvals[0]; oVals[2]=tvals[1];
                oVals[3]=uniMuVar[0]; oVals[4]=uniMuVar[1];
        }

     // Indicate when each thread finishes.
     //=====================================
     if (t1 == inparms[1] && (t2 == inparms[2] || inparms[1]==inparms[2]))
            cout << "Finishing thread #: " << inparms[5] << endl;
   }
}
```